\documentclass[
aps,
prd,
showpacs,
showkeys,
amsmath,
amssymb,
twocolumn,
floatfix,
superscriptaddress]{revtex4-2}

\usepackage{graphicx}
\usepackage{times}
\usepackage{verbatim}
\usepackage{color}
\usepackage{cancel}
\usepackage[normalem]{ulem}
\usepackage{bm}
\usepackage{url}
\usepackage{multirow}
\usepackage{textcomp}
\usepackage{dcolumn}
\usepackage{makecell}
\usepackage{booktabs}
\usepackage{siunitx}
\usepackage{xtab,afterpage}
\DeclareSIUnit{\charge}{\milli\volt\nano\second}
\DeclareSIUnit{\MeV}{\mega\electronvolt}
\DeclareSIUnit{\days}{days}
\usepackage{lineno}
\usepackage{enumerate}
\usepackage{appendix}
\usepackage{tabularx}
\usepackage[colorlinks,linkcolor=blue,anchorcolor=blue,citecolor=blue]{hyperref}
\usepackage{tikz}
\usetikzlibrary{shapes.geometric, arrows, backgrounds, mindmap, calc, positioning, intersections}
\usepackage{rotating}
\usepackage[version=3]{mhchem}
\usepackage{footnote}

\tikzset{
  startstop/.style = {rectangle, rounded corners, minimum width = 1cm, minimum height=0.5cm, text centered, draw = black},
  io/.style = {trapezium, trapezium left angle=70, trapezium right angle=110, minimum width=1cm, minimum height=0.5cm, text centered, draw=black},
  process/.style = {rectangle, minimum width=3cm, minimum height=1cm, text centered, draw=black},
  decision/.style = {diamond, aspect = 3, text centered, draw=black},
  arrow/.style = {->,>=stealth}
}

\definecolor{posurl}{cmyk}{.9 .9 0 0}
\hypersetup{
colorlinks=true
,urlcolor=posurl
,anchorcolor=posurl
,citecolor=posurl
,filecolor=posurl
,linkcolor=posurl
,menucolor=posurl
,linktocpage=true
,pdfa=true
}
\newcommand{\GEANT}{{\textsc{Geant}}}

\usepackage[caption=false]{subfig}

\begin{document}

\title{Mountain Muography for China Jinping Underground Laboratory}

\makeatletter
\newcommand{\fmarki}{*}
\newcommand{\fmarkii}{\ensuremath{\dagger}}
\newcommand{\fmarkiii}{\ensuremath{\ddagger}}
\newcommand{\fmarkiv}{\ensuremath{\mathsection}}
\newcommand{\fmarkv}{\ensuremath{\mathparagraph}}
\newcommand{\fmarkvi}{\ensuremath{\|}}
\newcommand{\fmarkvii}{**}
\newcommand{\fmarkviii}{\ensuremath{\dagger\dagger}}
\newcommand{\fmarkix}{\ensuremath{\ddagger\ddagger}}
\def\@fnsymbol#1{{\ifcase#1\or \fmarki\or \fmarkii\or \fmarkiii\or \fmarkiv\or \fmarkv\or \fmarkvi\or \fmarkvii\or \fmarkviii\or \fmarkix \else\@ctrerr\fi}}
\makeatother
\renewcommand{\fmarki}{\ensuremath{\dagger}}

\newcommand{\TUDEP}{\affiliation{Department of Engineering Physics \& Center for High Energy Physics, Tsinghua University, Beijing 100084, China}}
\newcommand{\TUPRI}{\affiliation{Key Laboratory of Particle \& Radiation Imaging (Tsinghua University), Ministry of Education, China}}
\newcommand{\HNUSPE}{\affiliation{School of Physics \& Electronics, Hunan University, Changsha 410082, China}}
\newcommand{\HNUPKL}{\affiliation{Hunan Provincial Key Laboratory of High-Energy Scale Physics and Applications,  Changsha 410082, China}}
\newcommand{\NKU}{\affiliation{School of Physics, Nankai University, Tianjin 300071, China}}
\newcommand{\UCASP}{\affiliation{School of Physical Sciences, University of Chinese Academy of Sciences, Beijing 100049, China}}
\newcommand{\SYSUP}{\affiliation{School of Physics, Sun Yat-Sen University, Guangzhou 510275, China}}
\newcommand{\NUSP}{\affiliation{School of Physics, Nanjing University, Nanjing 210093, China}}
\newcommand{\SDU}{\affiliation{Institute of Frontier and Interdisciplinary Science, Shandong  University, Qingdao, 266237, China}}
\newcommand{\JDUFS}{\affiliation{Jinping Deep Underground Frontier Science and Dark Matter Key Laboratory of Sichuan Province, China}}
\newcommand{\YRHDC}{\affiliation{Yalong River Hydropower Development Company, Ltd., 288 Shuanglin Road, Chengdu 610051, China}}
\newcommand{\LZU}{\affiliation{School of Nuclear Science and Technology \& MOE Frontiers Science Center for Rare Isotopes, Lanzhou University, Lanzhou 730000, China}}

\author{Xinshun Zhang}\TUDEP\TUPRI
\author{Shaomin Chen}\TUDEP\TUPRI
\author{Wei Dou}\TUDEP\TUPRI
\author{Haoyang Fu}\TUDEP\TUPRI
\author{Lei Guo}\TUDEP\TUPRI
\author{Ziyi Guo}\TUDEP\TUPRI
\author{XiangPan Ji}\NKU
\author{Jianmin Li}\TUDEP\TUPRI
\author{Jinjing Li}\email{Corresponding author: lijinjing@hnu.edu.cn}\HNUSPE
\author{Bo Liang}\TUDEP\TUPRI
\author{Ye Liang}\TUDEP\TUPRI
\author{Qian Liu}\UCASP
\author{Wentai Luo}\TUDEP\TUPRI
\author{Ming Qi}\NUSP
\author{Wenhui Shao}\TUDEP\TUPRI
\author{Haozhe Sun}\TUDEP\TUPRI
\author{Jian Tang}\SYSUP
\author{Yuyi Wang}\TUDEP\TUPRI
\author{Zhe Wang}\TUDEP\TUPRI
\author{Changxu Wei}\TUDEP\TUPRI
\author{Jun Weng}\TUDEP\TUPRI
\author{Yiyang Wu}\TUDEP\TUPRI
\author{Benda Xu}\TUDEP\TUPRI
\author{Chuang Xu}\TUDEP\TUPRI
\author{Tong Xu}\TUDEP\TUPRI
\author{Tao Xue}\TUDEP\TUPRI
\author{Haoyan Yang}\TUDEP\TUPRI
\author{Yuzi Yang}\LZU\TUDEP
\author{Aiqiang Zhang}\TUDEP\TUPRI
\author{Bin Zhang}\TUDEP\TUPRI
\author{Yang Zhang}\SDU
\author{Zhicai Zhang}\TUDEP\TUPRI
\author{Lin Zhao}\TUDEP\TUPRI
\author{Yangheng Zheng}\UCASP

\collaboration{JNE Collaboration}\noaffiliation
\date{\today}

\begin{abstract}
     The China Jinping Underground Laboratory (CJPL), located $\sim 2,400$~m beneath Jinping Mountain, is one of the world's deepest and largest ($\sim 300{,}000~\mathrm{m}^3$) underground facilities, hosting dark matter, nuclear astrophysics, and neutrino experiments. We report the first muon radiography (muography) conducted at this extraordinary depth. Cosmic muons detected by a one-ton prototype developed for the Jinping Neutrino Experiment were used to perform non-invasive subsurface density mapping over a 3~km lateral range. The 1.3~m diameter detector provides nearly isotropic acceptance and an angular resolution of $\sim 4.5^\circ$. By correlating the predicted surface muon flux distributions with the underground measurements, we reconstruct a directional opacity map that constrains the density structure of the overburden and shows excellent agreement with satellite-derived terrain models. This work demonstrates the feasibility of muography at extreme depths with kilometer-scale overburden and establishes a robust methodology for future geophysical applications and large-scale facilities, such as the full Jinping Neutrino Experiment. Based on this validated overburden model, we further predict the total muon fluxes for the eight experimental halls in CJPL-II, providing essential input for their physics programs.
\end{abstract}

\maketitle

\section{Introduction}
 Jinping Mountain in southwest China, located along the Yalong River, features a substantial elevation drop that enables the operation of multiple large hydroelectric power stations~\cite{shiyong2010jinping}. To support transportation and infrastructure development, two parallel traffic tunnels, each 17.5~km in length, have been excavated through the mountain. Situated between these tunnels, the China Jinping Underground Laboratory (CJPL) is currently the deepest underground research facility worldwide in terms of vertical rock overburden, and one of the largest in total volume~\cite{Cheng:2017usi}.

\begin{figure*}[!htbp]
    \centering
    \includegraphics[width=1.0\linewidth]{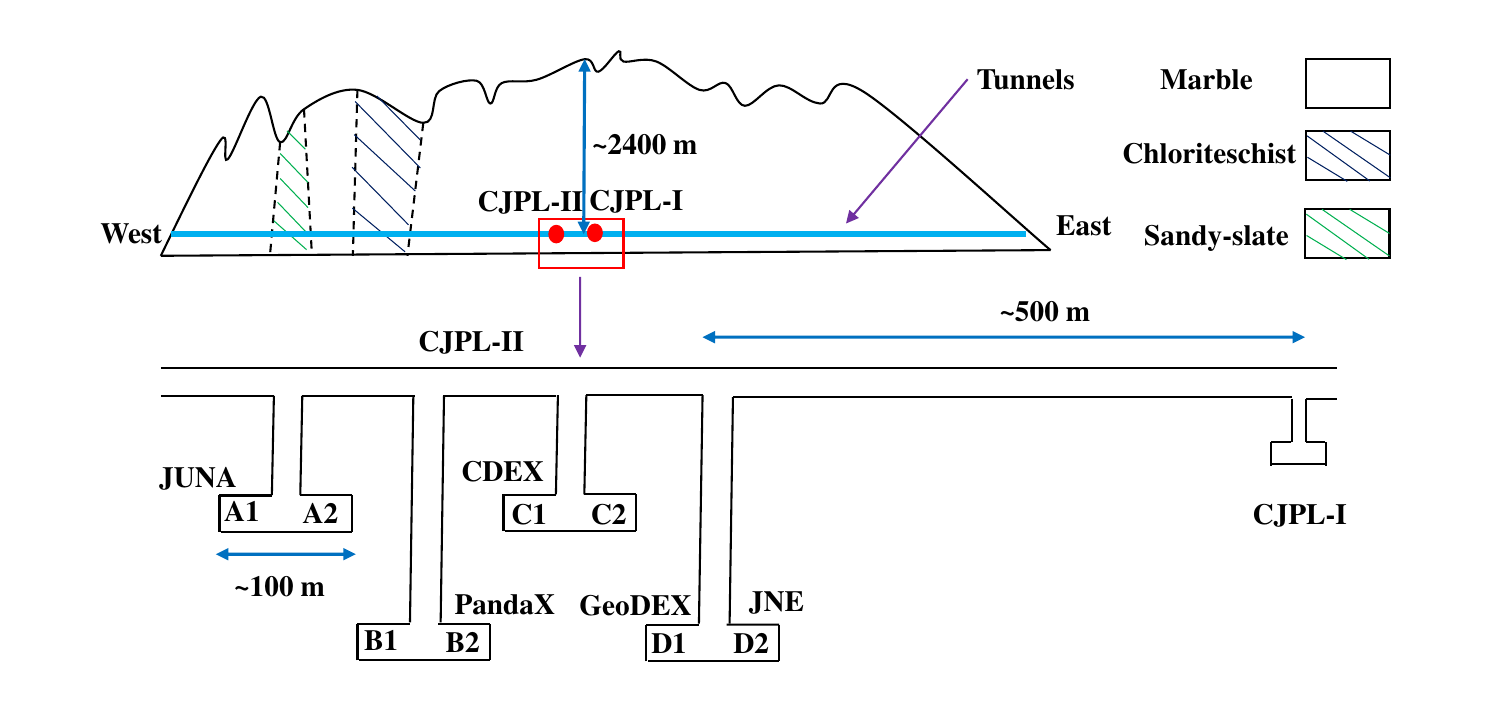}
    \caption{Schematic layout of the China Jinping Underground Laboratory (CJPL).}
    \label{fig:laboratory}
\end{figure*}

Taking advantage of its exceptional geological shielding, CJPL has been developed in two phases, CJPL-I and CJPL-II, each accommodating a broad range of low-background physics experiments. CJPL-I has hosted several small-scale detectors over the past decade, enabling key technical developments and initial physics measurements~\cite{CDEX:2022dda,PandaX-II:2021kai,Wu:2022oxo}. In CJPL-II, multiple large-volume experiments are currently operating or under construction, including the dark matter search experiments like CDEX and PandaX, the Jinping Neutrino Experiment (JNE), the Jinping Underground Nuclear Astrophysics facility (JUNA), and the underground geoscience project GeoDex~\cite{CDEX:2023vvc,PandaX-4T:2021bab,Jinping:2016iiq,JUNA:2022rvj,XIE20241}. In addition, several projects on neutrinoless double-beta decay, quantum measurement, ultra-pure material center, and low-background counting facility are under proposal or development~\cite{Chen:2022rzg}. The overall laboratory layout and corresponding rock overburden are shown in Fig.~\ref{fig:laboratory}. These characteristics establish CJPL as a uniquely valuable facility for rare-event searches and underground physics worldwide.

To support both scientific research and the safe operation of the laboratory, it is essential to obtain a comprehensive understanding of the internal structure of Jinping Mountain, which extends several kilometers laterally and reaches approximately 2000~m in elevation. Conventional geophysical imaging techniques, including electromagnetic, seismic, and gravitational methods, each face intrinsic limitations when applied to structures of this scale. Electromagnetic approaches are primarily sensitive to conductivity rather than density variations. Gravity surveys generally suffer from limited spatial resolution and non-unique inversion; and seismic methods rely on natural or artificial sources, which may introduce environmental disturbances~\cite{mao2023muon, shapiro2005high, barnoud2019bayesian}. In contrast, muography (cosmic muon radiography) provides a complementary and non-invasive imaging approach. Operating analogously to X-ray transmission but utilizing naturally occurring cosmic-ray muons, muography directly probes internal density distributions over large distances, while avoiding external excitation or structural interference~\cite{kaiser2019muography}. This feature makes muography particularly suitable for high-resolution, non-destructive imaging of large-scale geological formations such as mountain massifs.

Cosmic-ray muons are produced when high-energy primary cosmic rays interact with nuclei in the Earth’s atmosphere~\cite{Gaisser:2016uoy}. As muons propagate through matter, they lose energy and scatter through interactions with electrons and nuclei, thereby encoding information about the density and composition of the material they traverse. Owing to their relatively large mass and high kinetic energy, muons can penetrate kilometer-scale overburdens, which makes them an effective natural probe for imaging dense and extended structures~\cite{Nagamine:1995np,Borozdin:2003}. Muography has therefore been applied in a range of fields, including archaeology, volcanology, and large-scale geological studies, using detector technologies such as plastic scintillators, nuclear emulsions, and gas-based tracking systems~\cite{Han:2020ryi, tioukov2019first, Varga:2016axy}. However, for targets on the scale of entire mountains, these conventional detector systems face inherent limitations, particularly in angular acceptance and effective detection area, which restrict their ability to achieve high-resolution, wide-field imaging over kilometer-scale baselines.

In this work, we employ a one-ton liquid scintillator detector at CJPL to detect cosmic-ray muons and perform muography of Jinping Mountain. The detector provides nearly uniform angular acceptance and achieves high-precision muon direction reconstruction. Using the muon dataset accumulated since 2017, we apply the flux-attenuation method to infer the internal density structure of the mountain. By comparing the predicted surface muon flux distributions with the underground measurements, we obtain directional flux attenuation and reconstruct the mountain's density profile. In addition, the muographic results allow us to predict the underground muon fluxes in the eight experimental halls of CJPL-II. These predictions are particularly relevant for future experiments, especially in halls where full-coverage muon flux measurements are not feasible.

The remainder of this paper is organized as follows.
Sec.~\ref{sec:experiment} introduces the muography methodology and the muon detection system.
Sec.~\ref{sec:reconstruction} outlines the data processing framework, including energy reconstruction, event selection, and muon direction reconstruction.
Sec.~\ref{sec:location} details the determination of the detector’s location and orientation.
Sec.~\ref{sec:tomography} presents the muographic imaging results of the mountain.
Sec.~\ref{sec:prediction} provides predictions of muon fluxes in the CJPL-II halls, followed by conclusions and outlook in Sec.~\ref{sec:conclusion}.

\section{Methodology}\label{sec:experiment}

\subsection{Surface Muon Calculation}
We define the surface muon flux as the cosmic-ray muon flux at sea level. It is characterized by the three-dimensional distribution $\phi_s(E,\theta, \varphi)$, where $E$ is the muon energy, and $\theta$ and $\varphi$ are muon's zenith and azimuth angles, respectively. Here, $\theta$ is the angle from the vertical, and $\varphi$ is the angle from East in the horizontal plane. The CJPL facility is shielded by approximately 2,400~m of rock overburden. Consequently, muons reaching the underground laboratory must have energies on the TeV scale before entering the mountain. At such high energies, azimuthal asymmetry induced by geomagnetic and other effects, which is significant only at low energies ($\leq$100~GeV), can be neglected. The surface flux distribution is therefore assumed to be symmetric in $\varphi$ and simplifies to $\phi_s(E,\theta)$~\cite{Gaisser:2002jj}. Furthermore, these energetic muons originate in the upper atmosphere and have a negligible decay probability. Thus, any elevation-induced variations in $\phi_s(E,\theta)$ are also negligible. Although CJPL is located at an altitude of 1,600~m and Jinping Mountain's elevation is up to 4,000~m, the muon flux on the mountain surface can still be approximated as the sea level distributions. In this work, we adopt several surface muon flux distributions derived from different theoretical calculations, as described below.

Cosmic-ray muons are products of extensive air showers (EAS) initiated by the collision of a high-energy primary cosmic ray, of galactic or extragalactic origin, with an atmospheric nucleus~\cite{Gaisser:2016uoy}. This phenomenological process generates a large number of hadrons. The majority of muons arise from the decays of long-lived mesons ($\pi^\pm$ and $K^\pm$), while a small fraction also originates from the decays of short-lived hadrons, such as charmed ($D^\pm$) and unflavored ($\eta$) mesons~\cite{Illana:2010gh}. A standard empirical parameterization of the surface muon flux is Gaisser's formula developed using a semi-analytic method~\cite{Gaisser:2016uoy}. Accurate theoretical calculation of the flux requires modeling the hadronic interactions and particle production within the EAS, typically described by a set of coupled cascade equations. Several techniques have been developed to solve these equations, including full Monte-Carlo simulations and numerical methods~\cite{Gaisser:1983vc,Fedynitch:2015zma}. 

\begin{figure}[!htbp]
    \centering
    \includegraphics[width=0.95\linewidth]{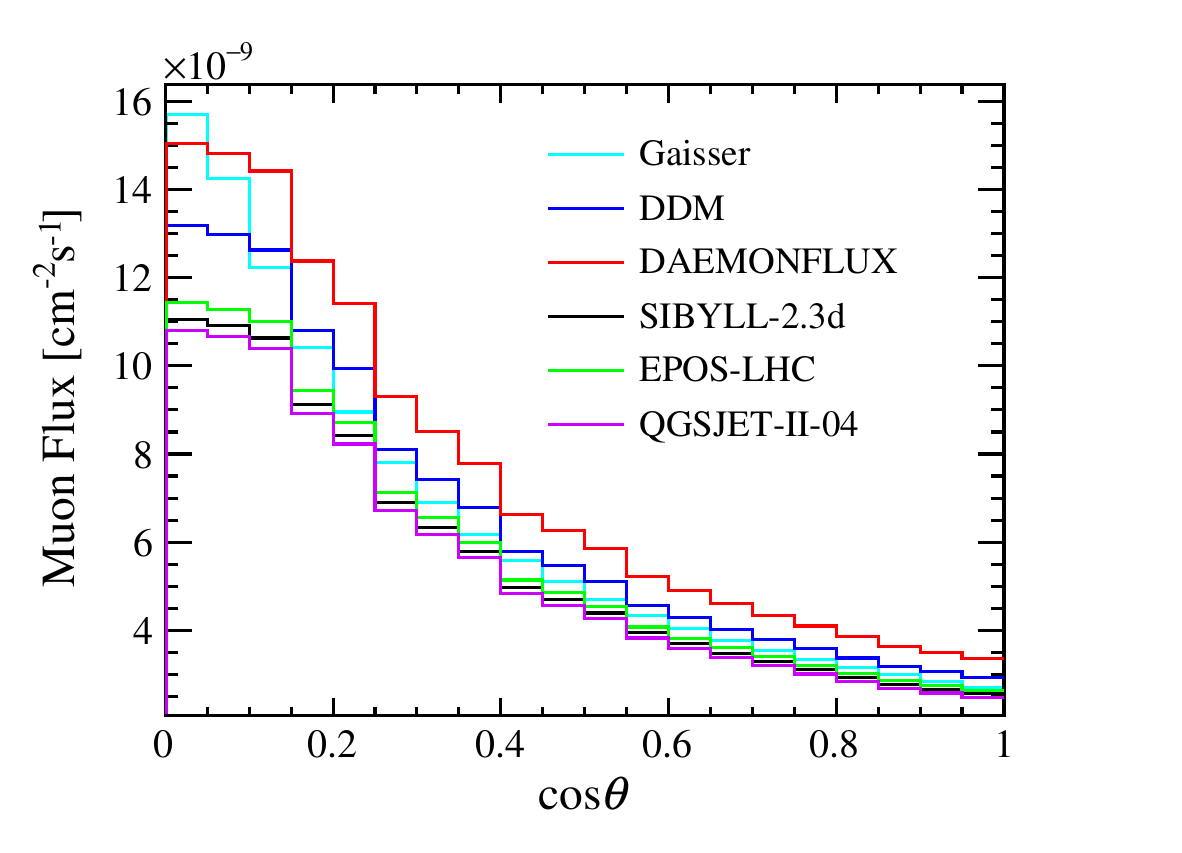}
    \caption{Zenith angle $\theta$ distribution of the surface flux for muons with energy exceeding 2~TeV. The fluxes are derived using the MCEq package with different hadronic models (solid lines) and fixed GSF primary spectra, compared to the parameterization from Gaisser's formula (dashed line).}
    \label{fig:zenith}
\end{figure}

In this study, we employ the Matrix Cascade Equations (MCEq) package, which provides numerical solutions to the cascade equations with computational speed superior to other methods while maintaining comparable accuracy~\cite{Fedynitch:2018cbl}. MCEq requires several inputs, including the primary cosmic ray energy spectra for all mass components and a hadronic interaction model. We use the Global Spline Fit (GSF) model, a state-of-the-art global fit to experimental cosmic-ray data with few theoretical assumptions, to describe the primary spectra~\cite{Dembinski:2017zsh}. Modeling the hadronic interactions in an EAS from first principles is challenging, as they lie in the non-perturbative QCD regime~\cite{Albrecht:2021cxw}. We therefore employ several phenomenological interaction models tuned to LHC data (post-LHC models). Among these, SIBYLL-2.3d and QGSJET-II-04 are developed specifically for EAS simulation, while EPOS-LHC, primarily designed for collider experiments, is also applicable~\cite{Riehn:2019jet,Pierog:2013ria,Ostapchenko:2010vb}. Additionally, other parameterization models, such as the Data-driven Model (DDM) and DAEMONFLUX, have been developed. These are driven by fixed-target accelerator data and subsequently calibrated against muon measurements~\cite{Fedynitch:2022vty,Yanez:2023lsy}. Figure~\ref{fig:zenith} shows the surface muon flux integrated over various zenith angle bins, calculated using MCEq with different hadronic models and the GSF primary spectrum, for muon energies exceeding 2~TeV.

\subsection{Muon Propagation in Rock}
Cosmic-ray muons lose kinetic energy in matter via ionization and radiative processes. Ionization predominates at low energies, whereas radiative processes, including bremsstrahlung, pair production, and photonuclear interactions, become dominant at energies above the muon critical energy ($\sim$700~GeV in standard rock)~\cite{Groom:2001kq}. After losing all kinetic energy, a muon either decays or is captured by a nucleus. Additionally, multiple Coulomb scattering with nuclei causes slight deflections in the muon trajectory~\cite{Scott:1963xw}. 

The rates of these interactions are directly related to the density and elemental composition of the medium, which is the fundamental principle of muography. For small-scale objects, muography is often performed using scattering, measuring the deviation between the incident and exit directions. This transmission-based approach is not suitable for large-scale structures such as mountains because of the significant absorption of the muon flux. Instead, muography of geological structures relies on measuring flux attenuation.

For a specific material, the survival probability $U(E_k, X)$ of a muon with initial energy $E_k$ after traversing a slant depth $X = \rho L$ (in units of m.w.e.) can be calculated theoretically or via simulation. This function is primarily determined by the elemental composition of the material, which governs the muon energy-loss rate, $dE/dX$. Per unit mass thickness, ionization loss scales as $Z/A$, while bremsstrahlung and pair production scale as $Z^2/A$~\cite{Groom:2001kq}. Photonuclear interactions have a more complex nuclear dependence~\cite{Woodley:2024eln}. The rock in Jinping Mountain is predominantly marble (CaCO$_3$), with primary elemental constituents of C, O, Mg, and Ca~\cite{CDEX:2021cll}. For comparison, Table~\ref{tab:rock} lists the compositions and effective $A$, $Z$, and $Z^2/A$ values for marble, standard rock, and crustal rock. The marble in Jinping exhibits a slightly higher energy-loss rate (a few percent) than the other rock types.

\begin{table*}[!htbp]
    \tabcolsep=0.28cm
    \centering
    \caption{Elemental composition (by mass fraction) for different rock types, along with their corresponding average atomic mass $A$, average atomic number $Z$, and $Z^2/A$. The Jinping Mountain rock is primarily marble. Elements with negligible content are omitted. Other rock types are also shown for comparison. Standard Rock is a hypothetical material defined by effective nuclear properties ($Z=11$, $A=22$, $Z/A=0.5$, $\rho=2.65$~g/cm$^3$) and does not correspond to any single element.}\label{tab:rock}
    \begin{tabularx}{\textwidth}{ccccccccccccc}
        \hline
        \hline
         Element & O & Si & Al & Fe & Ca & Na & K &Mg &C &$A$ &$Z$ &${Z^2}/{A}$\\
        \hline
         Standard & --& --&  --&  --&  --&  --& --& --& --& 22& 11& 5.5\\
         Marble &46.4\%& 0.2\%& 0.15\%&  0.1\%&  32.0\%&  0.01\%&  0.07\%& 11.5\%& 9.6\%& 24.31& 12.15&6.07 \\
         Crustal & 46.6\%& 27.7\%& 8.13\% & 5.0\% & 3.63\% & 2.83\% & 2.59\%& 2.09\%& 1.41\% &24& 11.83&5.83\\
        \hline
        \hline
    \end{tabularx}
\end{table*}

\begin{figure}[!htbp]
    \centering
    \includegraphics[width=0.95\linewidth]{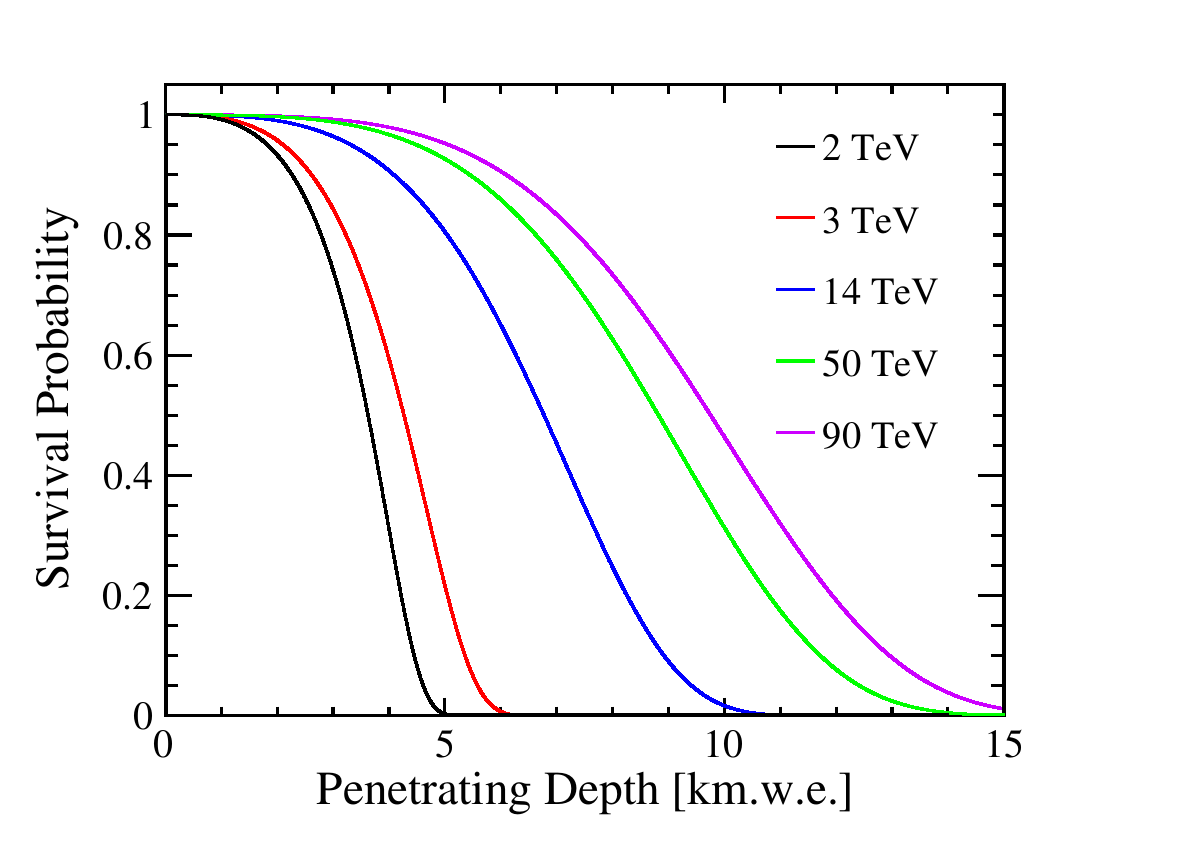}
    \caption{Muon survival probability $U(E_k, X)$ as a function of slant depth $X$ for various initial kinetic energies $E_k$ in Jinping marble, derived from \GEANT4 simulations~\cite{GEANT4:2002zbu,allison2006geant4}.}
    \label{fig:probability}
\end{figure}

Figure~\ref{fig:probability} shows the \GEANT4-based simulation of $U(E_k, X)$ for muons in Jinping marble~\cite{GEANT4:2002zbu,allison2006geant4}. The effective survival probability $P(X)$ for the entire muon spectrum is obtained by convolving $U(E_k, X)$ with the normalized surface muon energy distribution $f(E_k)$, evaluated at the corresponding zenith angle for each angular bin:
\begin{equation}
    P(X) = \int U(E_k, X) f(E_k) dE_k.
\end{equation}
Experimentally, we measure the underground muon flux $\phi(\theta,\varphi)$. The measured survival probability $P(\theta,\varphi)$ is then expressed as the ratio of the underground flux to the surface flux:
\begin{equation}
   P(\theta,\varphi) = \frac{\phi(\theta,\varphi)}{\phi_s(\theta,\varphi)},
\end{equation}
where $\phi_s(\theta,\varphi) = \int \phi_s(E,\theta,\varphi)\,dE$ is the energy-integrated surface flux.
By inverting the simulated $P(X)$ function, the muon slant depth in each direction, $X(\theta,\varphi)$, is inferred from the measured $P(\theta,\varphi)$. This directional map of $X$ reveals the mountain's internal density structure.

\subsection{Detection and Simulation}
\begin{figure}[!htbp] \centering \includegraphics[width=0.95\linewidth]{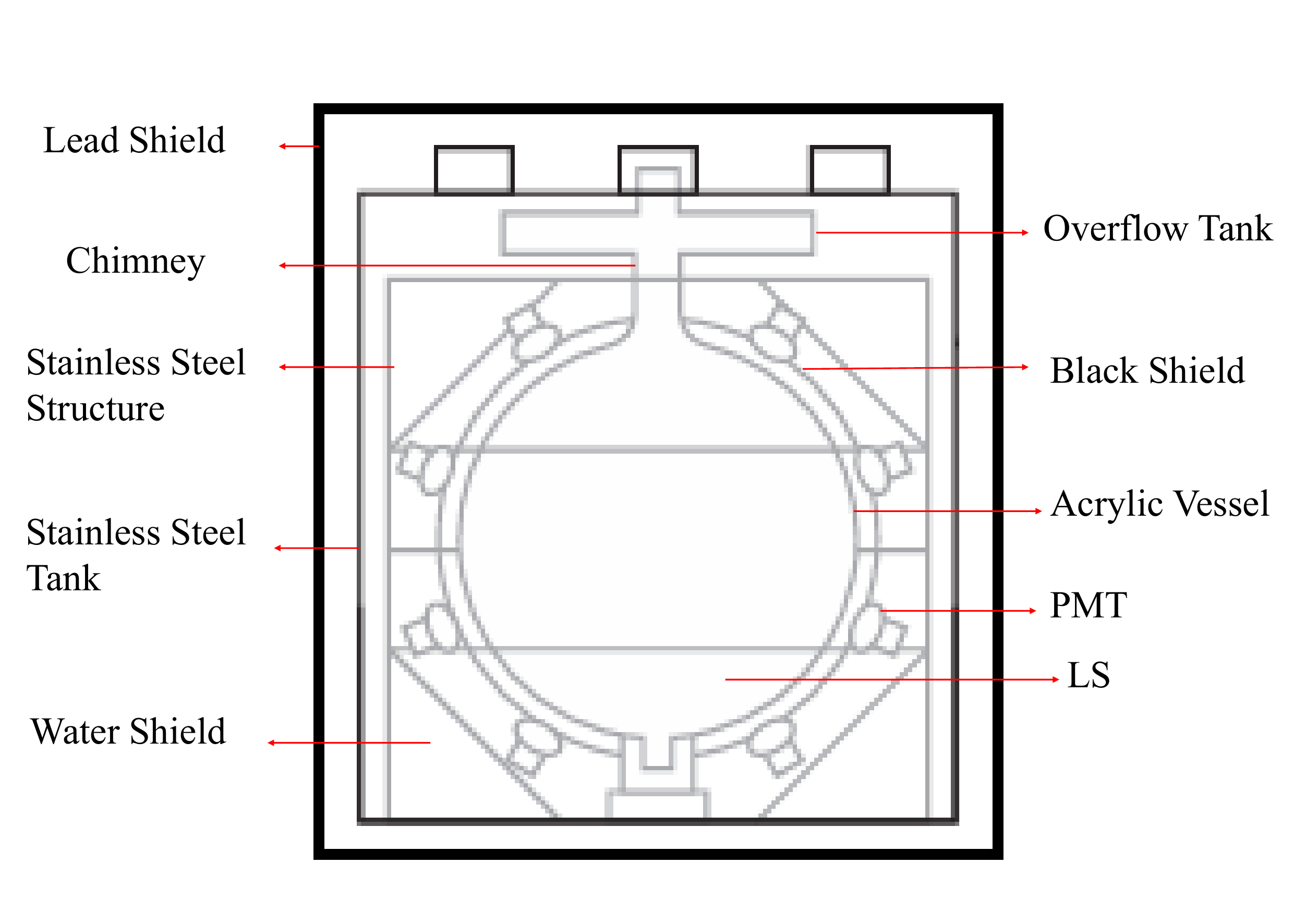} \caption{Schematic of the JNE 1-ton prototype detector.} \label{fig:detector} \end{figure}

Muon detection is performed using the JNE one-ton liquid scintillator prototype detector, which was constructed for technology validation and underground radioactivity measurements~\cite{Wu:2022oxo,JNE:2024gov}. The detector's primary component, as shown in Fig.~\ref{fig:detector}, is a 645-mm-radius spherical acrylic vessel with a 20-mm thickness that contains one ton of liquid scintillator (LS). The LS is composed of LAB as a solvent, doped with 0.07~g/L of the fluor 2,5-diphenyloxazole (PPO) and 13~mg/L of the wavelength shifter 1,4-bis(2-methylstyryl)benzene (bis-MSB).

Outside the vessel, thirty 8-inch Hamamatsu R5912 photomultiplier tubes (PMTs) are installed on a stainless steel structure, facing inward to detect photons produced inside the vessel. The PMTs are arranged uniformly in four levels: 5 each in the top and bottom levels, and 10 each in the two middle levels. This design provides a nearly uniform detection response over a 4$\pi$ solid angle. A black shield outside the acrylic vessel blocks photons produced outside the LS, ensuring that all photons received by the PMTs originate from within the LS.

When a muon traverses the detector, it deposits energy in the LS, producing numerous scintillation photons. These photons propagate to the PMTs, which detect and then convert the light into electrical signals. A high-precision electronic system, consisting of four CAEN V1751 FlashADC boards and a CAEN V1495 logical trigger module, records the timing and charge information of these signals. Each FlashADC board has eight channels, 10-bit ADC precision for a 1~V dynamic range, and a 1~GHz sampling rate. This system collects data as digitized pulse waveforms, facilitating the subsequent reconstruction of the muon's trajectory.

The detector assembly (vessel, PMTs, and stainless steel structure) is housed within a stainless steel tank filled with purified water. The tank is then surrounded by a 5-cm-thick lead wall. These structures combine to provide passive shielding against ambient natural radiation.

We developed a \GEANT4-based simulation framework to model the detector's response to muon signals~\cite{GEANT4:2002zbu,allison2006geant4}. The simulation incorporates the complete detector geometry, including a one-meter-thick layer of surrounding rock to account for accompanying shower particles (primarily electromagnetic) produced by muons interacting near the detector. Muon energy loss and subsequent interactions are handled by the standard electromagnetic and QGSP\_BERT hadronic physics lists. The framework also models the full optical physics, including the production of scintillation and Cherenkov photons, their propagation, and their detection by PMTs, as well as the electronic readout response. This setup ensures that the simulation generates events in the same format as the experimental data, enabling identical reconstruction procedures. Further details of the detector simulation can be found in Refs.~\cite{JNE:2020bwn,JNE:2024gov}.

\section{Data Analysis}\label{sec:reconstruction}
The data set used in this analysis was collected from July 31, 2017, to March 27, 2024, corresponding to an effective live time of 1338.6 days after data-quality checks. The data-quality is monitered using the waveform features in the data, while about 5\% of the data with large electronic fluctuations and abnormal PMT occupancies are excluded.  This dataset represents an increase of 160.6 days compared to the data set used in Ref.~\cite{JNE:2024gov}. The same data set is employed in Ref.~\cite{Zhang:2025tmz}.

\subsection{Energy Reconstruction}
The deposited energy of each event is reconstructed from the total charge recorded by the PMTs. Initially, PMTs exhibiting low occupancy or high electronic noise are identified as ``bad channels'' and excluded from the reconstruction. The gain of each remaining PMT is calibrated using intrinsic dark noise signals, converting the integrated waveform charge into the number of photoelectrons (PEs). The sum of PEs across all functional PMTs serves as the primary measure of reconstructed energy. Given the detector's small volume, corrections for position-dependent non-uniformity are negligible. The energy scale is then calibrated by converting the PE sum to physical units (e.g., MeV) using well-defined radioactive sources, as detailed in Refs.~\cite{Wu:2022oxo,JNE:2024gov}.

\begin{figure}[!htbp]
\centering
    \subfloat[Desaturation]{\includegraphics[width=0.95\linewidth]{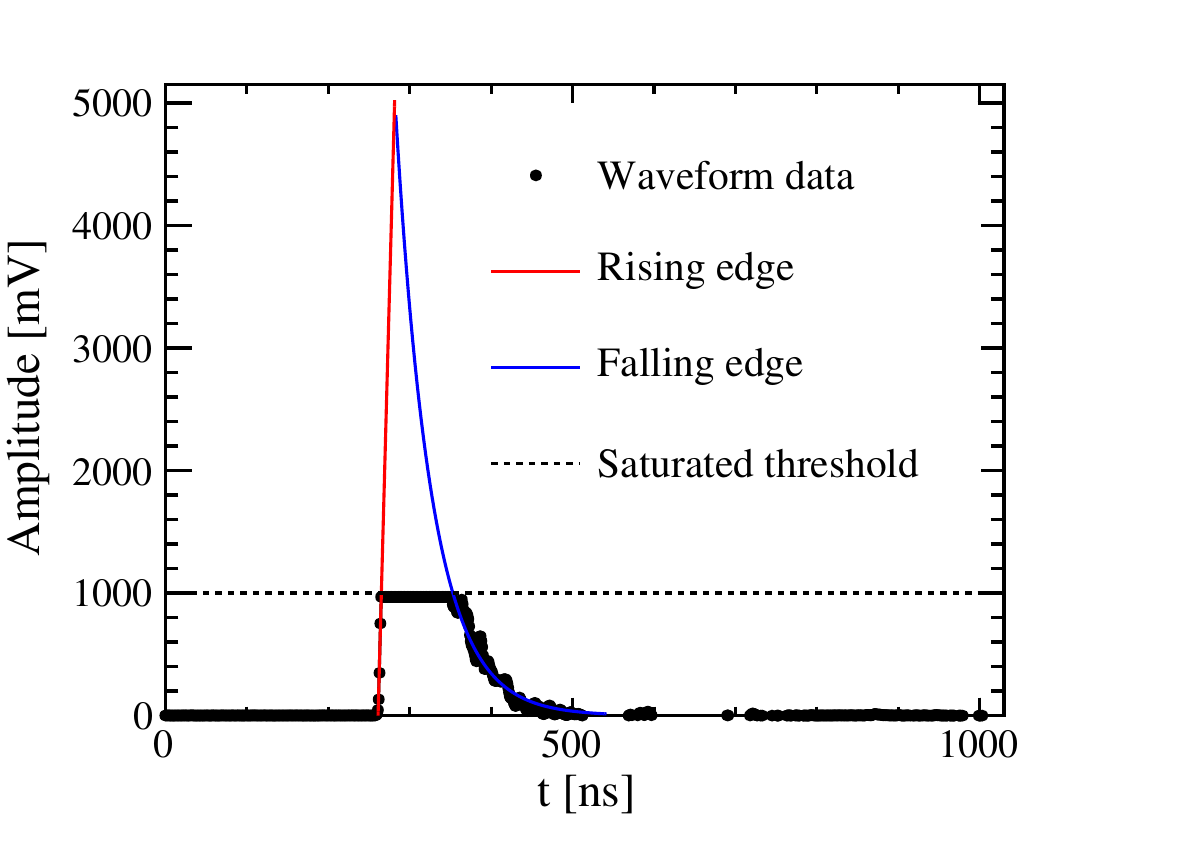}\label{fig:waveform}}\\
    \subfloat[Performance]{\includegraphics[width=0.95\linewidth]{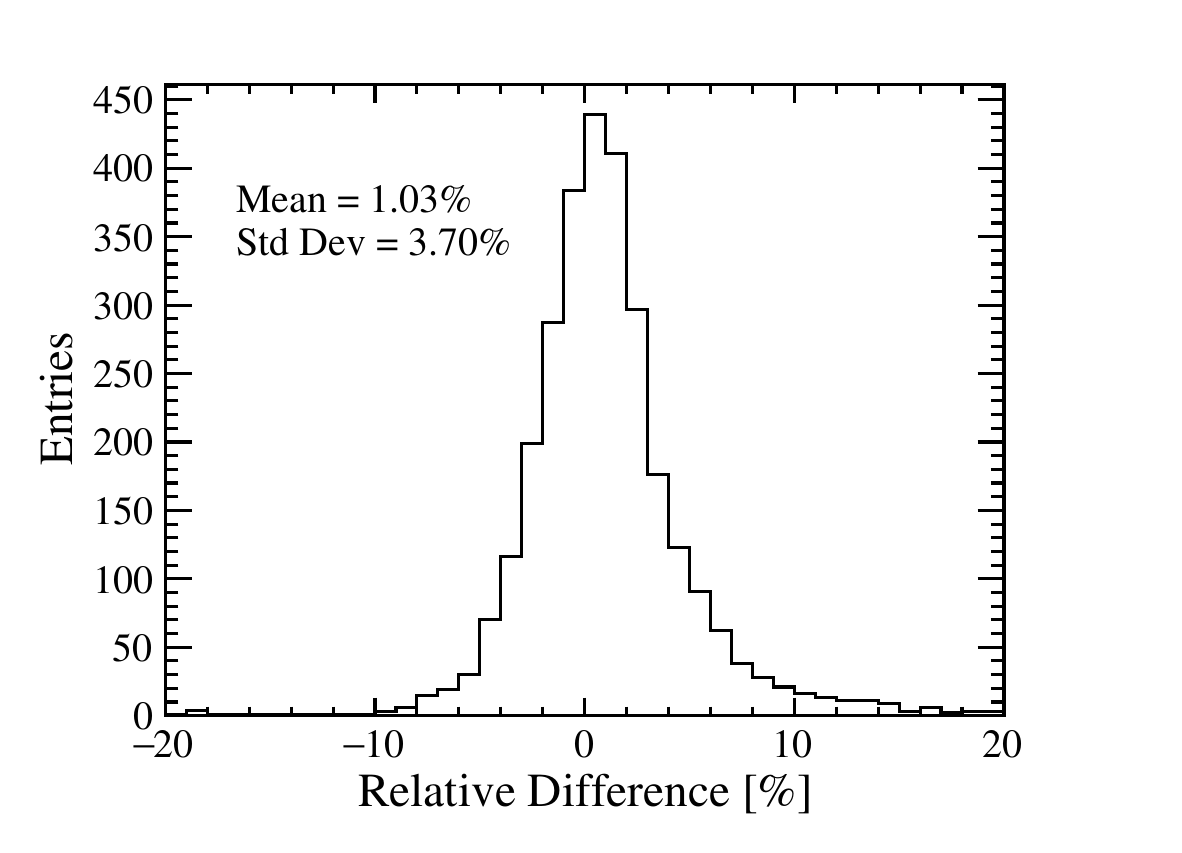}\label{fig:resolution}}
    \caption{(a): Schematic of the desaturation algorithm. The rising (linear) and falling (exponential) edges are fit and extrapolated to reconstruct the saturated portion of the waveform. (b): Performance of the algorithm on an unsaturated test sample. The relative difference is defined as $(C_{d}-C_{t})/C_{t}$, where $C_{d}$ and $C_{t}$ are the de-saturated and true charges, respectively.}
    \label{fig:saturation}
\end{figure}

High-energy signals, particularly from muons, often cause waveform saturation in the electronics due to their limited dynamic range. This saturation leads to an underestimation of the deposited energy. To mitigate this, a desaturation algorithm was developed. As illustrated in Fig.~\ref{fig:waveform}, the algorithm fits the non-saturated rising and falling edges of the waveform (using linear and exponential functions, respectively) and extrapolates these fits to reconstruct the charge in the saturated region. The total charge obtained from this desaturation procedure is then used for the energy reconstruction. The algorithm's performance was validated using a sample of unsaturated waveforms from the data, for which the true charge is known. For this test, the saturation threshold was artificially lowered, and the desaturation algorithm was applied to the resulting ``artificially saturated'' signals. The results demonstrate strong linearity between the desaturated and true charge. As shown in Fig.~\ref{fig:resolution}, the systematic bias originating from the method itself is estimated to be $\sim$1\%, with a resolution of $\sim$3.7\%.

\subsection{Event Selection}
The collected data consists primarily of muon events and backgrounds from natural radioactivity (e.g., uranium and thorium decay chains). To isolate muon signals, we require the deposited energy to exceed 90~MeV. This threshold effectively removes radioactive backgrounds, which are typically at the MeV scale. Remaining instrumental backgrounds, such as PMT light emission (flashers) and electronic noise, are rejected using two dedicated variables based on PMT charge patterns and waveform structure, as detailed in Refs.~\cite{Wu:2022oxo,JNE:2024gov}.

\begin{figure}[!htbp]
    \centering
    \includegraphics[width=0.95\linewidth]{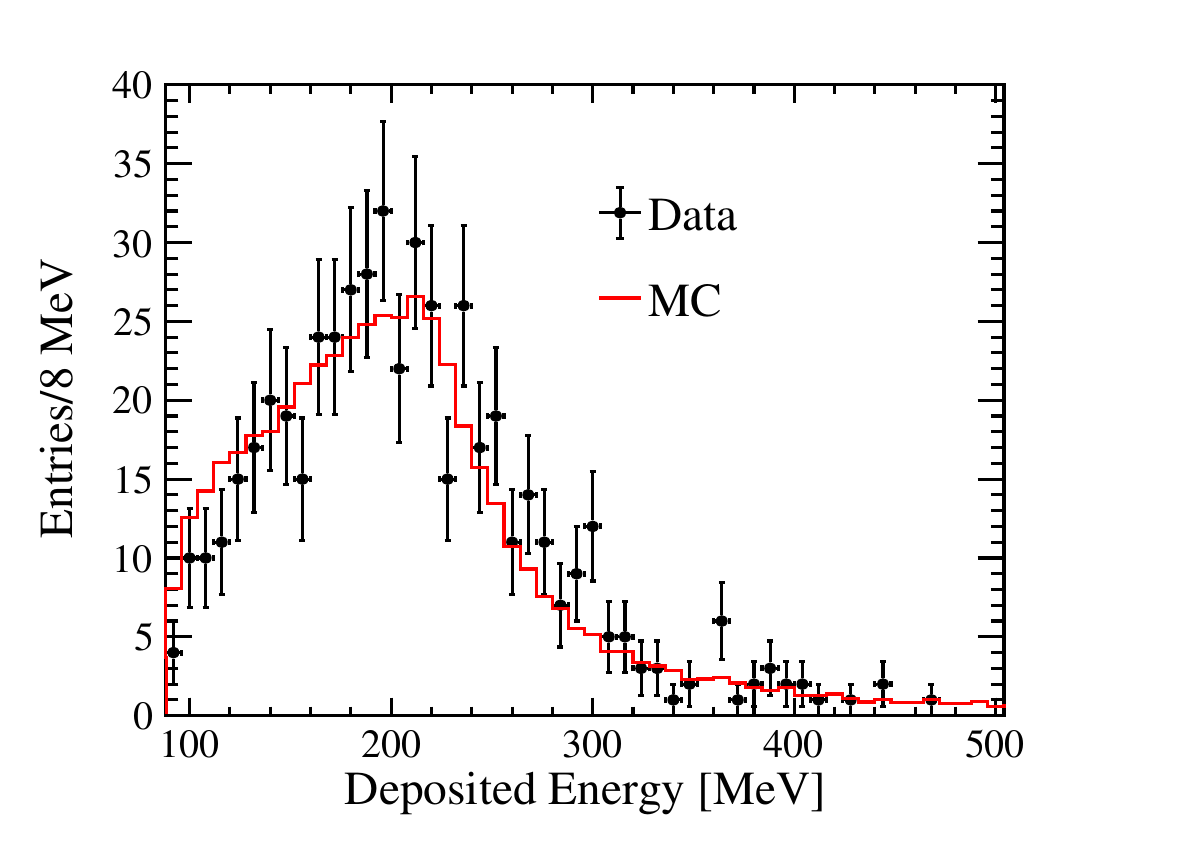}
    \caption{Deposited energy distribution of selected muon events from data (points) compared to the \GEANT4 simulation (histogram).}
    \label{fig:muonenergy}
\end{figure}

After all cuts, 547 muon events are selected for analysis. The deposited energy distribution of these events is consistent with simulation, as shown in Fig.~\ref{fig:muonenergy}. The muon selection efficiency, evaluated via simulation, is 82.2\%. Further simulation studies confirmed that this efficiency is uniform across the full 4$\pi$ solid angle, indicating an isotropic detector acceptance.

\subsection{Direction Reconstruction}
The direction of each selected muon event is reconstructed using an improved template-matching algorithm. This method, well-suited for small-volume detectors, compares the detector response of a data event to a large library of simulated muon templates. The direction is estimated based on the most similar templates. This work enhances previous iterations of the algorithm~\cite{JNE:2020bwn, JNE:2024gov} by incorporating both charge and timing information from each PMT.

The templates were generated using the full detector simulation. Muons were sampled uniformly over the 4$\pi$ solid angle, with entry points sampled across a hemisphere to ensure they traversed the target volume. Each template is tagged with its true direction and entry point. To ensure consistency, the same reconstruction and event selection criteria applied to the data were applied to the simulated events. This process yielded a final library of approximately 400,000 muon templates.

For each event (in data or simulation), we construct charge and time vectors, $\boldsymbol{C} = (c_0, c_1, \cdots, c_{29})$ and $\boldsymbol{T} = (t_0, t_1, \cdots, t_{29})$, where $c_i$ and $t_i$ are the calibrated charge and hit time of the $i$-th PMT, respectively. These vectors are then standardized:
\begin{equation}
    \boldsymbol{\Tilde{T}} = ((t_0-\mu_t)/\sigma_t, (t_1-\mu_t)/\sigma_t, \cdots, (t_{29}-\mu_t)/\sigma_t),
\end{equation}
\begin{equation}
    \boldsymbol{\Tilde{C}} = ((c_0-\mu_c)/\sigma_c, (c_1-\mu_c)/\sigma_c, \cdots, (c_{29}-\mu_c)/\sigma_c),
\end{equation}
where $\mu_c$ ($\mu_t$) and $\sigma_c$ ($\sigma_t$) are the mean and standard deviation of the elements in $\boldsymbol{C}$ ($\boldsymbol{T}$), respectively. The final detector response vector, $\boldsymbol{R}$, is the concatenation of these two normalized vectors:
\begin{equation}
    \boldsymbol{R} = (\boldsymbol{\Tilde{T}}, \boldsymbol{\Tilde{C}}).
\end{equation}

Let $\boldsymbol{R_{d}}$ be the response vector for a data event, and $\boldsymbol{R_{t}}$ be the vector for a template. The similarity between them is quantified by the Euclidean distance $d$:
\begin{equation}
    d = \left| \boldsymbol{R_{d}} - \boldsymbol{R_{t}} \right|.
\end{equation}
For a given data event, the $k$-most similar templates (i.e., those with the smallest $d$) are identified. The final reconstructed direction, $\boldsymbol{P}$, is the weighted average of the directions $\boldsymbol{P_{i}}$ of these $k$ templates, where the weight is the reciprocal of the distance $d_i$:
\begin{equation}
    \boldsymbol{P} = \frac{\sum_{i=1}^{k} \frac{1}{d_{i}}\boldsymbol{P_{i}}}{\sum_{i=1}^{k} \frac{1}{d_{i}}},
\end{equation} 
where $\boldsymbol{P_{i}}$ is the unit direction vector of the $i$-th template muon and $d_{i}$ is its distance to the event being reconstructed. The resulting vector $\boldsymbol{P}$ is then normalized to unit length before extracting $(\theta, \varphi)$.

\begin{figure}[!htbp]
    \centering
    \includegraphics[width=0.95\linewidth]{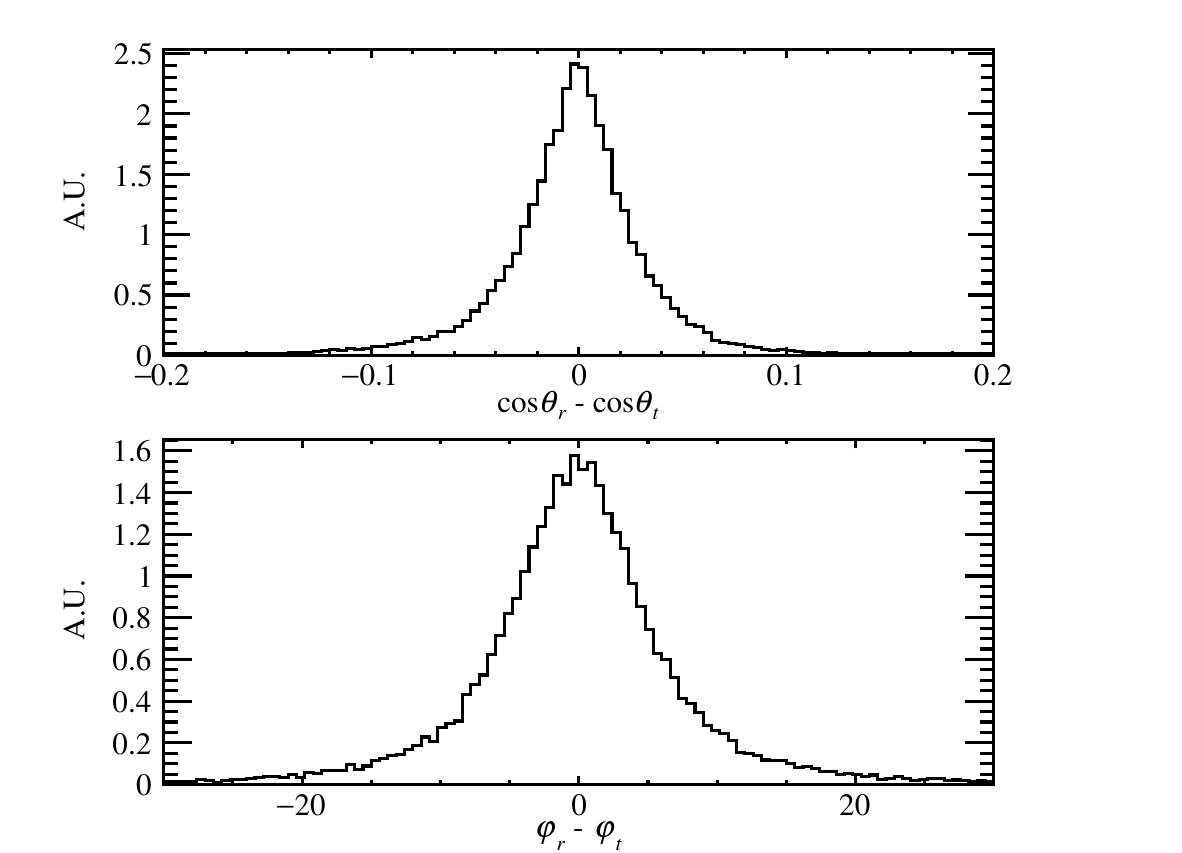}
    \caption{Reconstruction performance evaluated with the simulation. The panels show the difference between reconstructed ($r$) and true ($t$) directions for the zenith ($\theta$) and azimuth ($\varphi$) angles. The reconstruction is unbiased, with a mean angular resolution of 4.5$^{\circ}$.}
    \label{fig:recon}
\end{figure}

A separate, statistically independent simulation sample was generated to optimize the parameter $k$ and evaluate the reconstruction performance. The angular resolution was found to stabilize for $k \geq 20$; we therefore adopt $k=20$ for all subsequent analysis. Using this optimized method, the mean angular separation (opening angle) between the true and reconstructed muon directions is 4.5$^{\circ}$, which represents a significant improvement over the 5.0$^{\circ}$ resolution achieved by the previous charge-only method~\cite{zhang2022mou}. As shown in Fig.~\ref{fig:recon}, the reconstruction performance is uniform across the full 4$\pi$ solid angle, and the biases in both zenith ($\theta$) and azimuth ($\varphi$) are negligible. These features are crucial for the subsequent angular flux analysis.

\section{Location and Orientation of the Detector}\label{sec:location}
Muon direction is parameterized by its zenith ($\theta$) and azimuth ($\varphi$) angles. While the high-energy (TeV-scale) surface muon flux is azimuthally symmetric, the mountain's topography is not. Therefore, the $\varphi$ distribution of muons measured underground, after penetrating the overburden, is highly sensitive to the surrounding terrain and, consequently, to the precise location of the laboratory within the mountain.

This sensitivity allows us to determine the detector's position. We simulated the expected muon azimuth distributions for a 100-meter grid of potential detector locations (latitude and longitude) centered on the assumed CJPL-I position. Furthermore, we scanned the detector's internal orientation (the rotation of its local x-axis relative to True East) over 360$^{\circ}$. The optimal position and orientation are found by minimizing a binned $\chi^2$ statistic, which quantifies the goodness-of-fit between the simulated and measured $\varphi$ distributions:
\begin{equation}
    \chi^2 = \sum_i \frac{\left[n^i_\mathrm{data}-n^i_\mathrm{sim}(\eta)\right]^2}{n^i_\mathrm{data}},
\end{equation}
where $n^i_\mathrm{data}$ and $n^i_\mathrm{sim}$ are the event counts in the $i$-th $\varphi$ bin for data and simulation, respectively, and $\eta$ represents the detector's orientation shift. This Neyman $\chi^2$ form, with data counts in the denominator, is adopted so that the variance estimate does not depend on the simulation normalization. Interpolation was used on the discrete grid points to create a continuous $\chi^2$ landscape.

\begin{figure}[!htbp]
    \centering
    \includegraphics[width=0.95\linewidth]{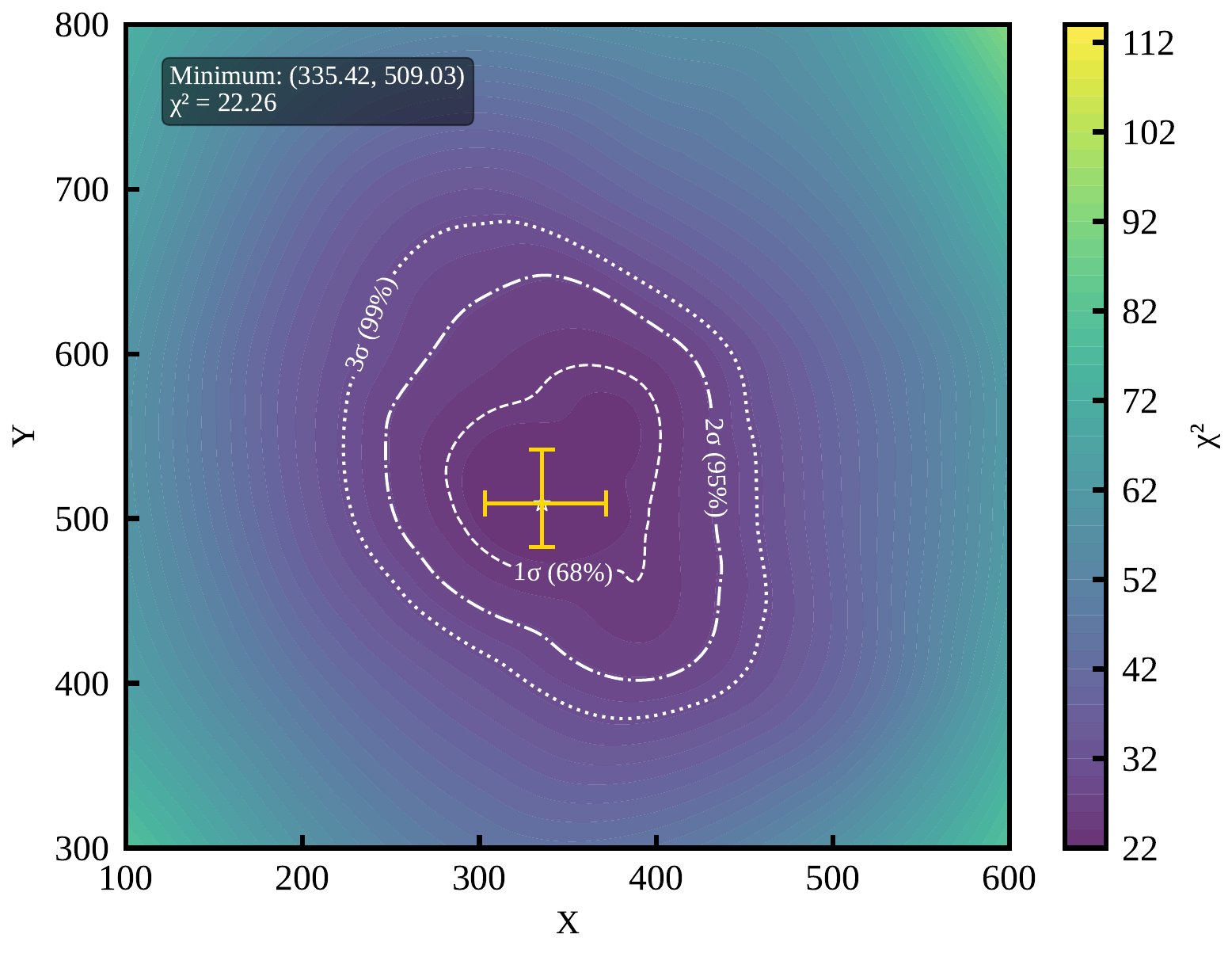}
    \caption{The 2-dimensional $\chi^2$ distribution as a function of the detector's relative position. $X$ and $Y$ are the relative longitude and latitude, respectively. The 1$\sigma$, 2$\sigma$, and 3$\sigma$ confidence regions are shown.}
    \label{fig:position}
\end{figure}

\begin{figure}[!htbp]
    \centering
    \includegraphics[width=0.95\linewidth]{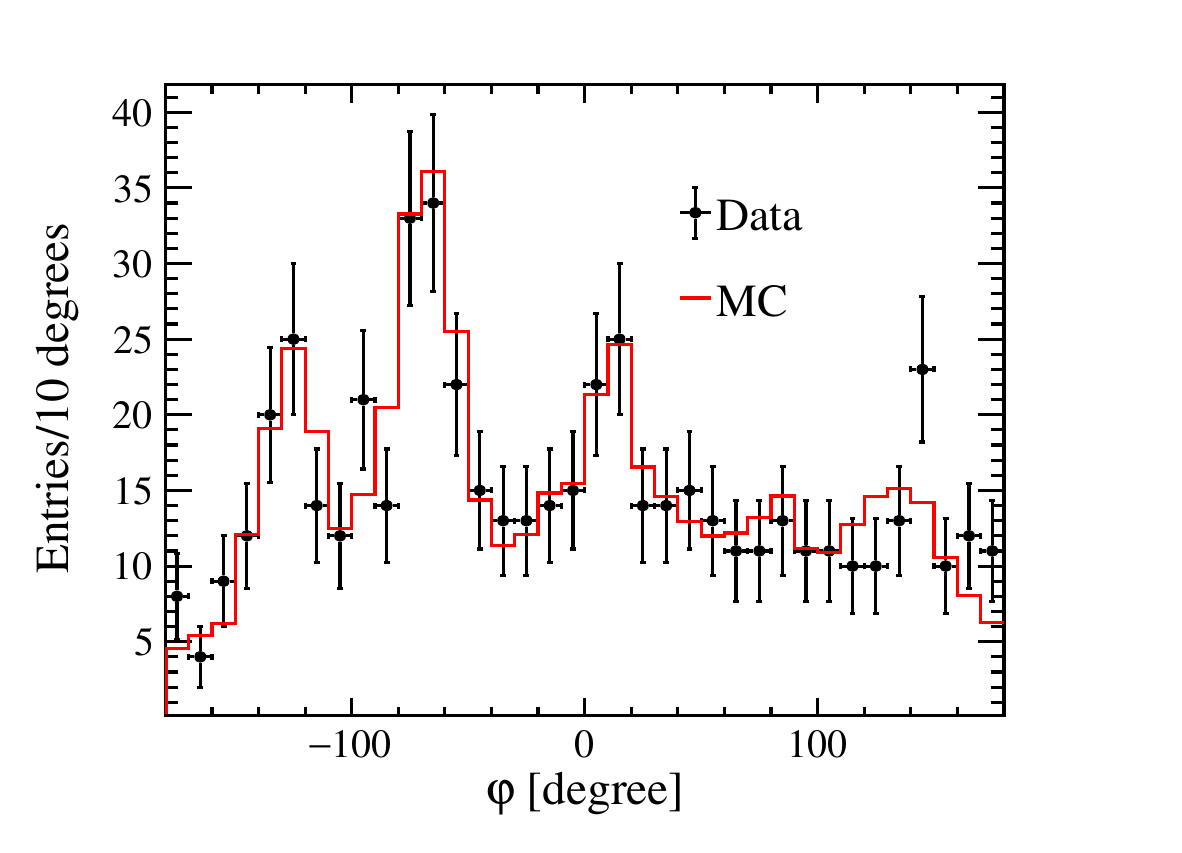}
    \caption{The azimuthal angle ($\varphi$) distribution for data (points) and the best-fit simulation (histogram). $\varphi=0$ corresponds to the True East direction.}
    \label{fig:phi}
\end{figure}

Figure~\ref{fig:position} shows the 2-dimensional $\chi^2$ distribution as a function of the detector's relative longitude and latitude. The best-fit position is found at $(335.4 \pm 36.4, 509.3 \pm 32.8)$~m relative to the initially assumed coordinates. The fit also determines that the detector's internal x-axis is rotated by $49.3 \pm 0.2$ relative to True East. The minimum $\chi^2$ is 22.3 for 33 degrees of freedom ($\chi^2/\mathrm{ndf}=0.67$), indicating good agreement. The $\varphi$ distributions for data and the best-fit simulation are shown in Fig.~\ref{fig:phi}.

\section{Muography}\label{sec:tomography}

\begin{figure*}[!htbp]
    \centering
    \subfloat[Measured Flux (cm$^{-2}$s$^{-1}$)]{\includegraphics[width=0.48\linewidth]{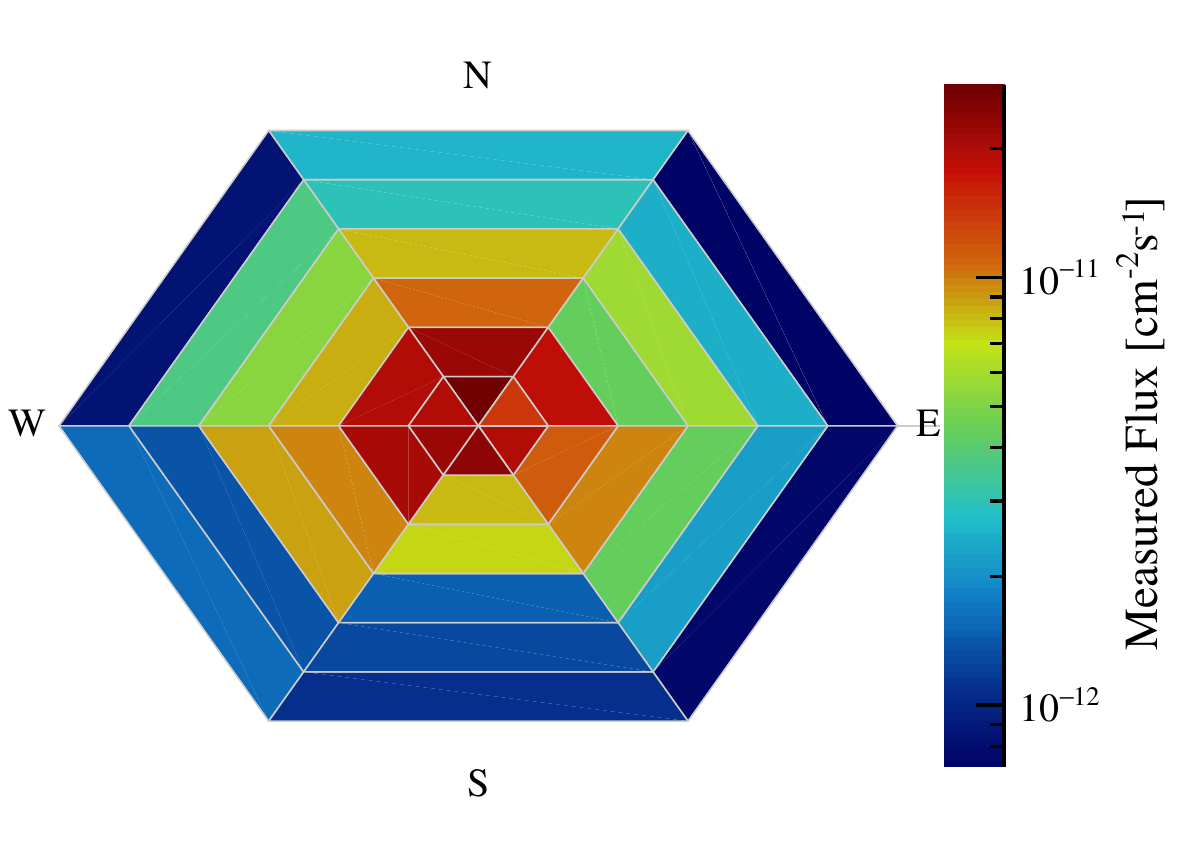}\label{fig:flux}}
    \subfloat[Measured Attenuation]{\includegraphics[width=0.48\linewidth]{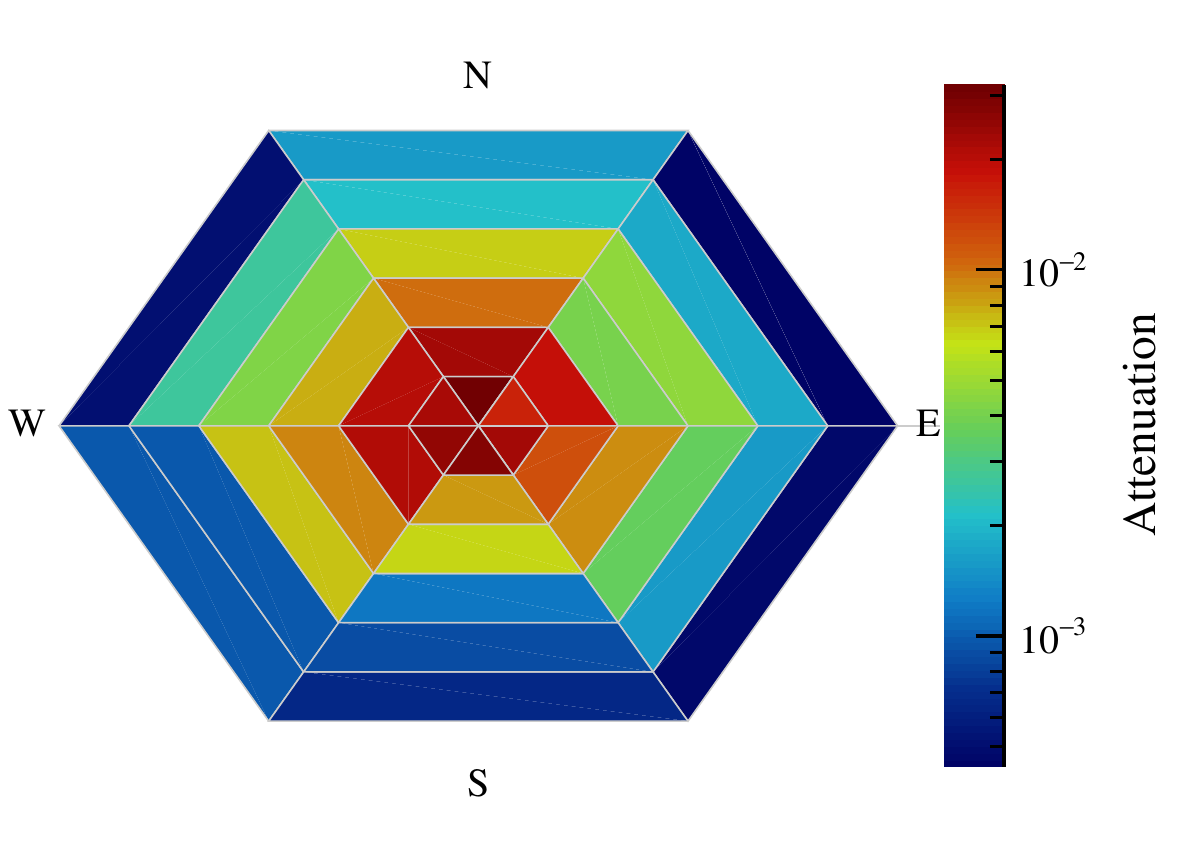}\label{fig:sp}}\\
    \subfloat[Measured Slant Depth (km.w.e.)]{\includegraphics[width=0.48\linewidth]{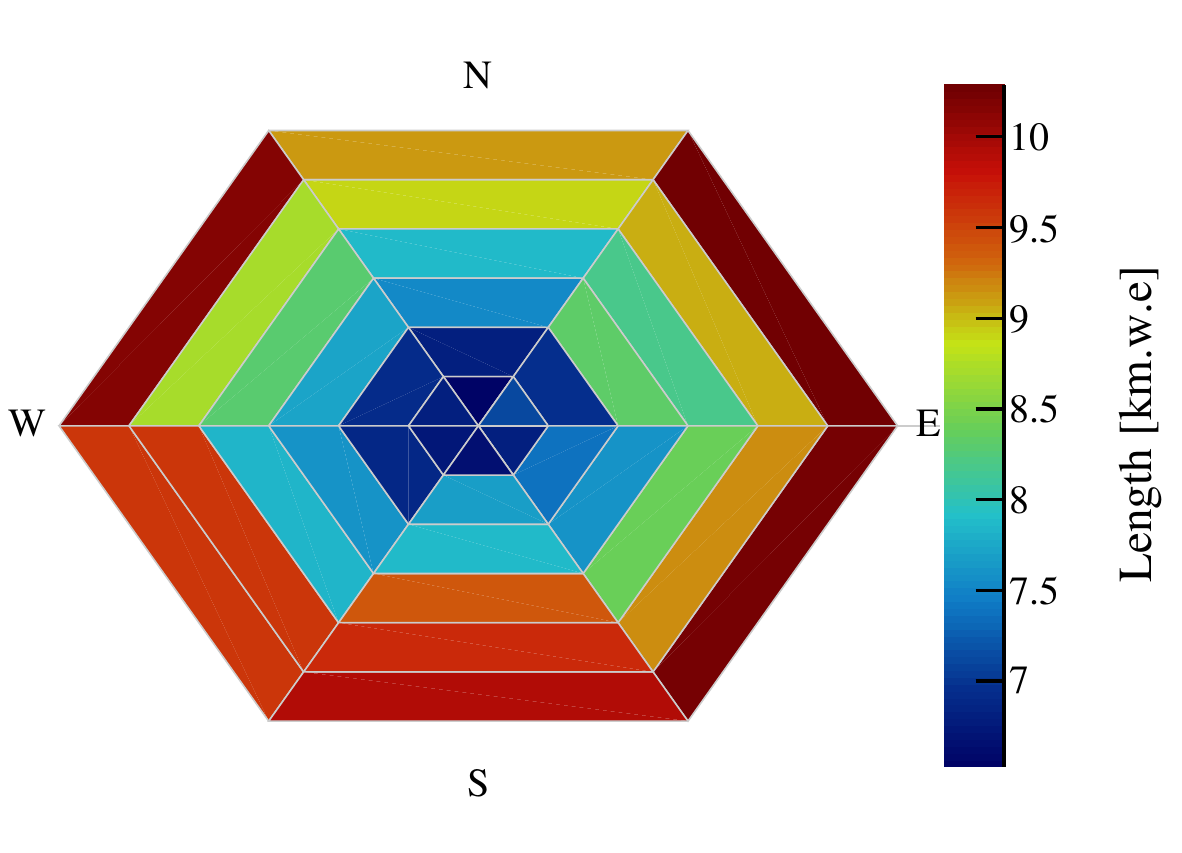}\label{fig:length}}
    \subfloat[Reconstructed Terrain]{\includegraphics[width=0.48\linewidth]{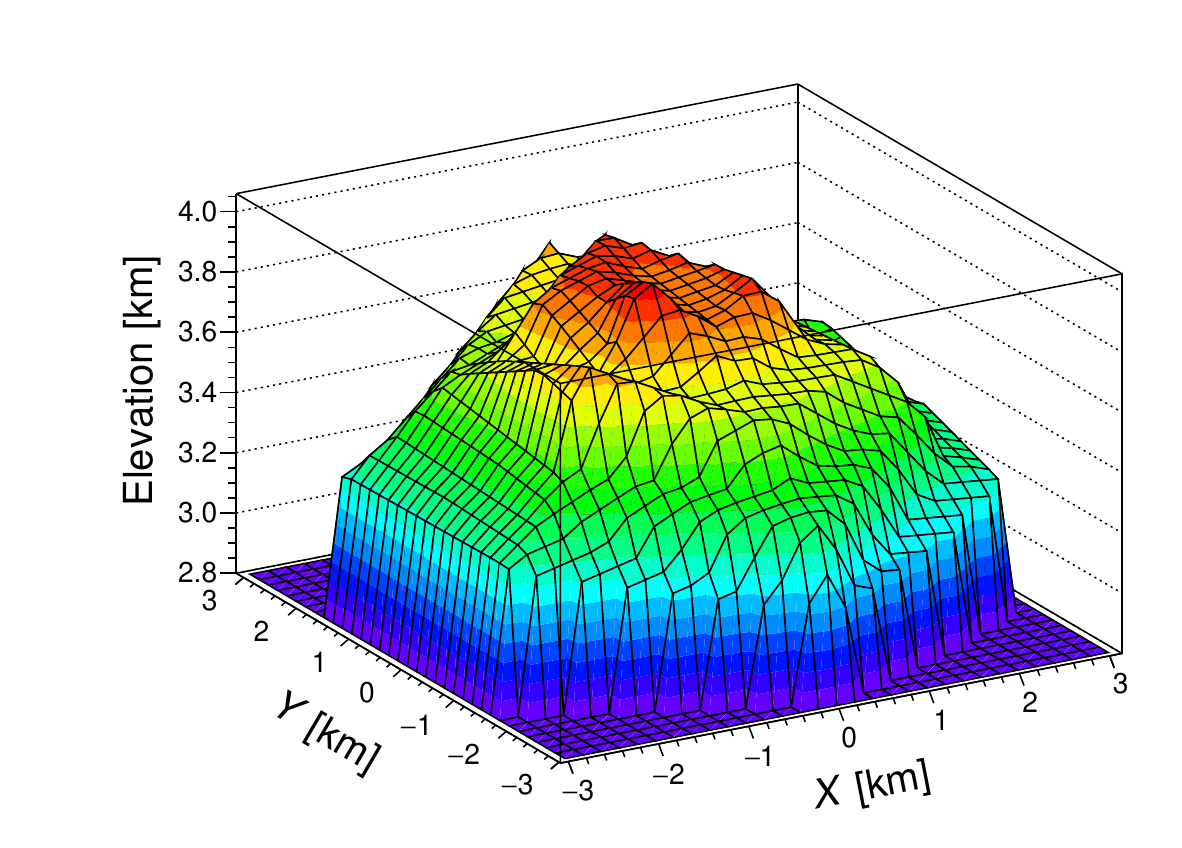}\label{fig:terrain}}
    \caption{(a): The measured underground muon flux $\phi(\theta,\varphi)$. (b): The measured flux attenuation $P(\theta,\varphi)$, derived using the nominal SIBYLL-2.3d surface flux. (c): The reconstructed slant depth $X_\mathrm{meas}(\theta,\varphi)$ (in km.w.e.) inferred from the attenuation map. (d): A 3D representation of the mountain's structure based on the measured slant depths, where the nominal density is applied here to convert the slant depth to elevation in units of kilometers. For the first three figures, the radial dimension is $\cos\theta$: the innermost point and the outermost circle correspond to the vertical ($\cos\theta=1$) and near-horizontal ($\cos\theta=0.4$) directions of muons, respectively. In the last figure, the axes $X$ and $Y$ represent relative longitude and latitude, respectively, with the coordinate origin at the position of CJPL-I.}
    \label{fig:muography}
\end{figure*}

Muography is performed by measuring the angular-dependent muon flux. We divide the angular phase space into a $6\times6$ grid in $\cos\theta$ and $\varphi$ for the region $\cos\theta>0.4$. Events with $\cos\theta<0.4$ (near-horizontal) are excluded due to limited statistics. The integrated muon flux in each angular bin $(\theta,\varphi)$ is:
\begin{equation}
    \phi(\theta,\varphi) = \frac{N(\theta,\varphi)}{\epsilon \cdot S(\theta,\varphi) \cdot T},
\end{equation}
where $N(\theta,\varphi)$ is the number of reconstructed muons in the bin, $T$ is the total live time, $\epsilon$ is the angular-independent selection efficiency (82.2\%), and $S(\theta,\varphi)$ is the detector's effective area for that direction, obtained from the detector simulation. The definitions of $\theta$ and $\varphi$ are consistent with previous sections. The resulting measured muon flux map is shown in Fig.~\ref{fig:flux}.

\begin{figure}[!htbp]
    \centering
    \includegraphics[width=0.95\linewidth]{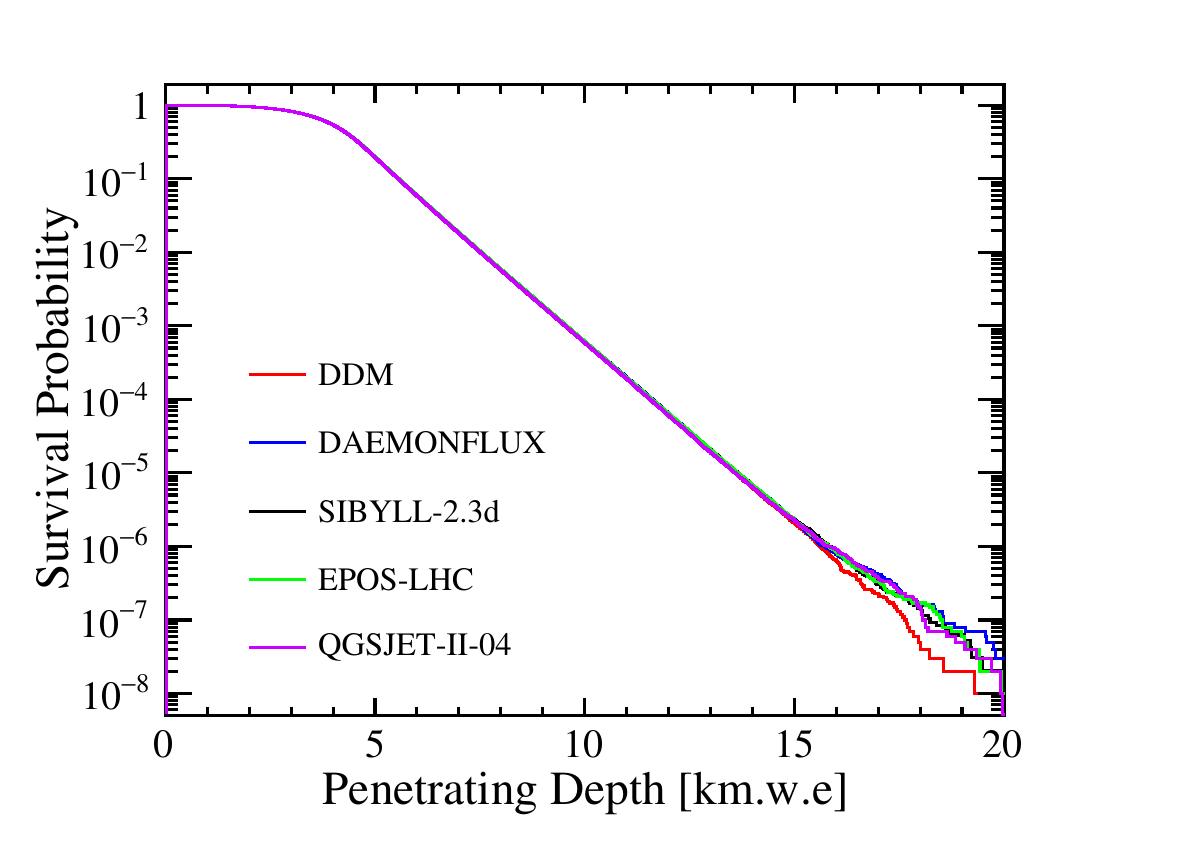}
    \caption{The effective survival probability profile $P(X)$ calculated using surface muon spectra from different hadronic models. The model dependence is negligible.}
    \label{fig:survival_model}
\end{figure}

Next, we derive the effective survival probability profile $P(X)$ by convolving the \GEANT4-simulated $U(E_k, X)$ (Fig.~\ref{fig:probability}) with the MCEq-calculated surface muon energy spectrum $f(E_k)$, as described in Sec.~\ref{sec:experiment}. Although the median minimum surface energy for muons to reach CJPL is $\sim$3~TeV, we set the lower integration boundary at 2~TeV to fully account for stochastic energy loss processes. For this profile, we use the SIBYLL-2.3d hadronic model as our nominal case. We evaluated the systematic uncertainty from this choice by recalculating $P(X)$ using other hadronic models (Fig.~\ref{fig:survival_model}). The differences in the resulting $P(X)$ are negligible for slant depths up to 15~km.w.e. This demonstrates that the shape of the survival probability curve is robust and largely independent of the specific hadronic interaction model.

To determine the flux attenuation, we must compare the underground flux $\phi(\theta,\varphi)$ (Fig.~\ref{fig:flux}) with the surface flux $\phi_s(\theta,\varphi)$. Unlike the $P(X)$ profile, the absolute normalization of the surface flux $\phi_s$ is highly model-dependent, varying by up to about 30\% between models. Consequently, the final muography result is intrinsically dependent on the choice of surface flux model. However, in the present study, both the reconstruction of the mountain profile and the search for localized density anomalies rely only on the relative directional observables and are therefore insensitive to the overall normalization. We  adopt SIBYLL-2.3d as the nominal model for all subsequent calculations.

The measured flux attenuation $P(\theta,\varphi) = \phi(\theta,\varphi) / \phi_s(\theta,\varphi)$ for each angular bin is shown in Fig.~\ref{fig:sp}. By inverting the $P(X)$ function (Fig.~\ref{fig:survival_model}), we convert each attenuation value $P(\theta,\varphi)$ into a measured slant depth, $X_\mathrm{meas}(\theta,\varphi)$, in units of km.w.e. The resulting map of measured slant depths is shown in Fig.~\ref{fig:length}. When projected, this 2D angular map constrains the density structure of the mountain, as depicted in Fig.~\ref{fig:terrain}.

The angular resolution of this muography method is primarily determined by the statistics of measured muon events rather than the detector's intrinsic resolution. In this work, due to the large overburden depth at CJPL, the muon flux is attenuated to an extremely low level, resulting in poor muography resolution for Jinping Mountain. We further evaluate the performance of muography with this detector at shallower depths, where the muon flux is significantly higher than at CJPL, thereby yielding a better muography resolution. Using the parameterized formula in Ref.~\cite{JNE:2020bwn}, we can calculate the muon fluxes and corresponding measured event rates at different overburden depths. The muography resolution is approximated as 
\begin{equation}
    \sigma = \sqrt{\frac{A \times B \times N}{R_{\mu} \times T}};
\end{equation}
where $A$ and $B$ represent the angular bin widths in $\theta$ and $\varphi$, respectively, $N$ is the total number of angular bins, $R_\mu$ is the detected muon event rate, and $T$ is the total data-taking time. Here $\sigma$ represents the angular bin size required to achieve a given statistical precision.

\begin{figure}[!htbp]
    \centering
    \includegraphics[width=0.95\linewidth]{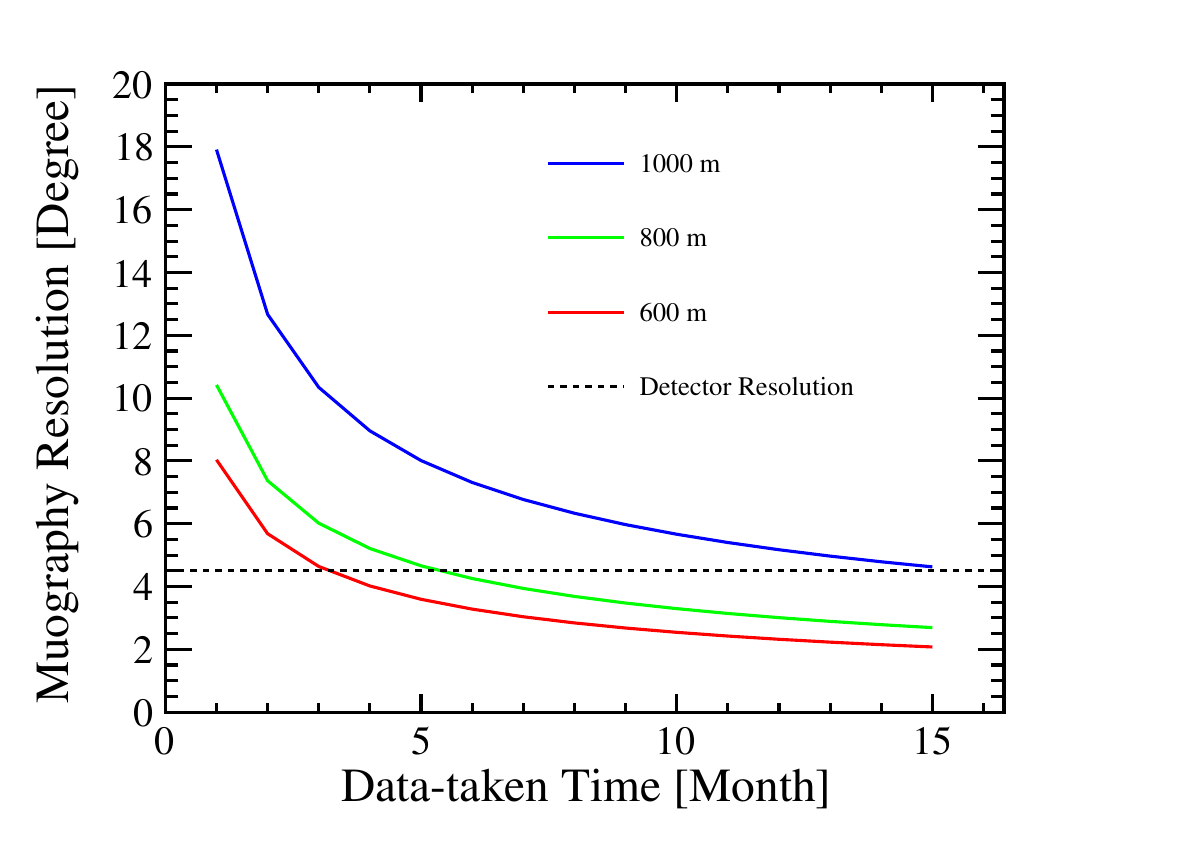}
    \caption{The evaluated performance of the muography method at various overburden depths. The detector's intrinsic resolution is also plotted here. The statistical uncertainties of measured fluxes in each angular bin are set to be less than 10\%.}
    \label{fig:depth}
\end{figure}

Figure~\ref{fig:depth} shows the calculated results at various depths, requiring that the statistical uncertainty in each angular bin remain below 10\%. For a 600~m overburden, the detector reaches the best muography resolution within approximately 3~months, at which point it becomes limited by the detector's intrinsic angular resolution of $4.5^{\circ}$. Greater overburden depths require longer data-taking periods to reach this intrinsic limit due to the lower muon flux.

These estimates establish two practical limits for this detector. The statistical-precision requirement (10\% per bin) corresponds to resolving a density anomaly region on the scale of 100~m.w.e., while the intrinsic angular resolution corresponds to a 70~m.w.e.\ scale region located 300~m from the detector. In summary, this muography method is effectively applicable to imaging density anomalies of 100~m.w.e.\ scale in mountains with heights up to 600~m.

To search for density anomalies, we compare our measured slant depth $X_\mathrm{meas}(\theta,\varphi)$ with an expected slant depth $X_\mathrm{exp}(\theta,\varphi)$. The expected value is calculated using the geographical path length $L_\mathrm{geo}(\theta,\varphi)$ (in meters) for each direction, derived from the NASA SRTM3 dataset~\cite{terrain2007} and our best-fit detector location. We then assume a nominal rock density $\rho_\mathrm{nom}$ (in g/cm$^3$, based on the marble composition in Table~\ref{tab:rock}) to convert this to m.w.e. We define the density ratio $R(\theta,\varphi)$ as:
\begin{equation}
    R(\theta,\varphi) = \frac{X_\mathrm{meas}(\theta,\varphi)}{X_\mathrm{exp}(\theta,\varphi)} = \frac{X_\mathrm{meas}(\theta,\varphi)}{\rho_\mathrm{nom} \times L_\mathrm{geo}(\theta,\varphi)} \approx \frac{\rho_\mathrm{avg}(\theta,\varphi)}{\rho_\mathrm{nom}}.
\end{equation}
This ratio, $R$, directly compares the average measured density, $\rho_\mathrm{avg}$, along a given path to the nominal rock density. If $R=1$, the measured density matches the nominal value. Deviations from $R=1$ would indicate density anomalies, such as cavities ($R<1$) or high-density ore veins ($R>1$). 

We construct mountain geometries with density anomalies to validate this method. Using the NASA SRTM3 satellite dataset, we construct the Jinping mountain terrain around CJPL in the simulation~\cite{terrain2007}. Then, the rock with a density of $2.8~\mathrm{g/cm}^3$ is filled inside the mountain, while a cube with a side length of 200~m is constructed and located 1000~m above the detector and 500~m to the East, which is filled with low density (air and water) or high density (iron ore with an assumed bulk density of $5.6~\mathrm{g/cm}^3$) matter. Muons are generated on the mountain surface in the simulation using the nominal mountain and different mountain geometries with density anomalies, while the muons reaching CJPL are recorded for further analysis. 

\begin{figure}[!htbp]
    \centering
    \includegraphics[width=0.95\linewidth]{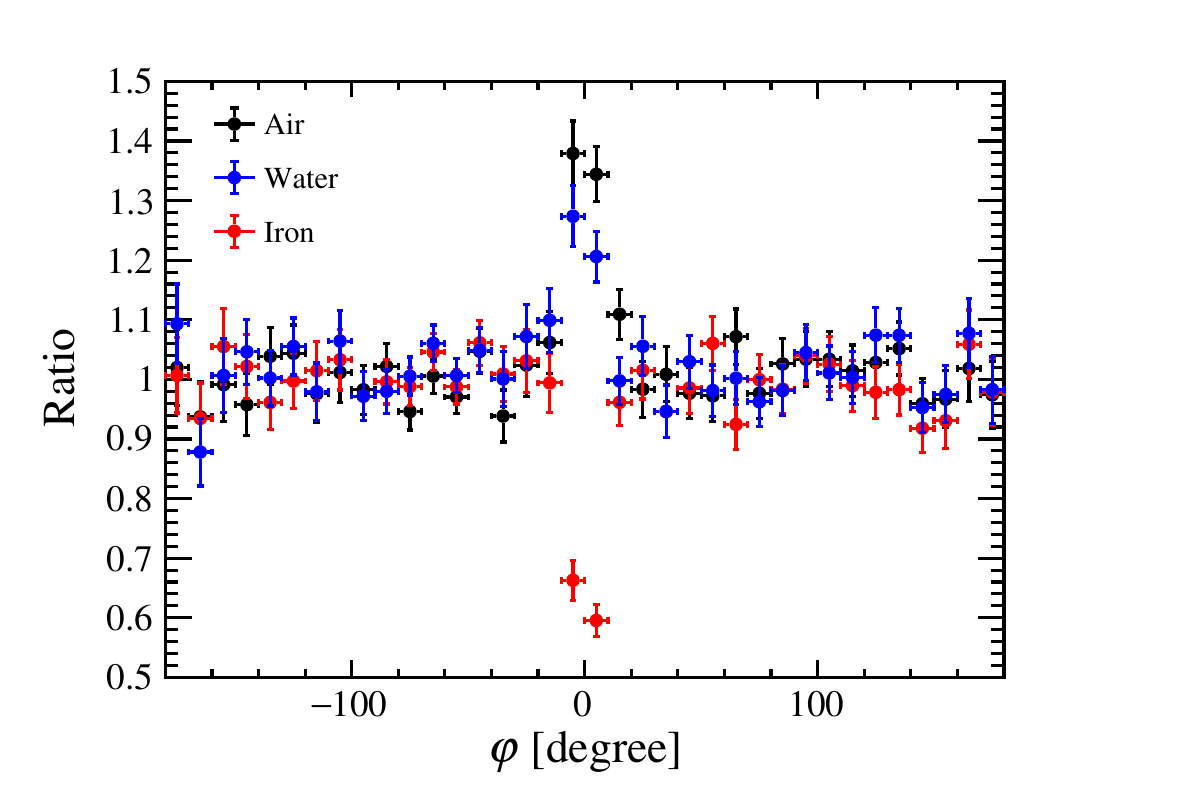}
    \caption{Comparison of the azimuth distributions for underground muons ($\cos\theta>0.8$) in the simulation between the nominal mountain geometry and those with a cube inside the mountain filled with air, water, and iron. $\varphi = 0^\circ$ corresponds to the True East direction.}
    \label{fig:compare}
\end{figure}

Figure~\ref{fig:compare} shows comparisons of the azimuth distributions for underground muons under different mountain geometries in the simulation. As expected, muons from regions with density anomalies (roughly $\cos\theta>0.8$ and $\lvert \varphi \rvert<22^\circ$) show significant deviations compared with other normal directions. For water and air, a smaller energy loss is expected in this direction, thus a larger muon intensity. Instead, for iron, the increased opacity along the line of sight results in greater muon attenuation, corresponding to a smaller survival probability and muon intensity. These simulation results sufficiently validate our method for identifying density anomalies in mountain structures.

\begin{figure}[!htbp]
    \centering
    \includegraphics[width=0.95\linewidth]{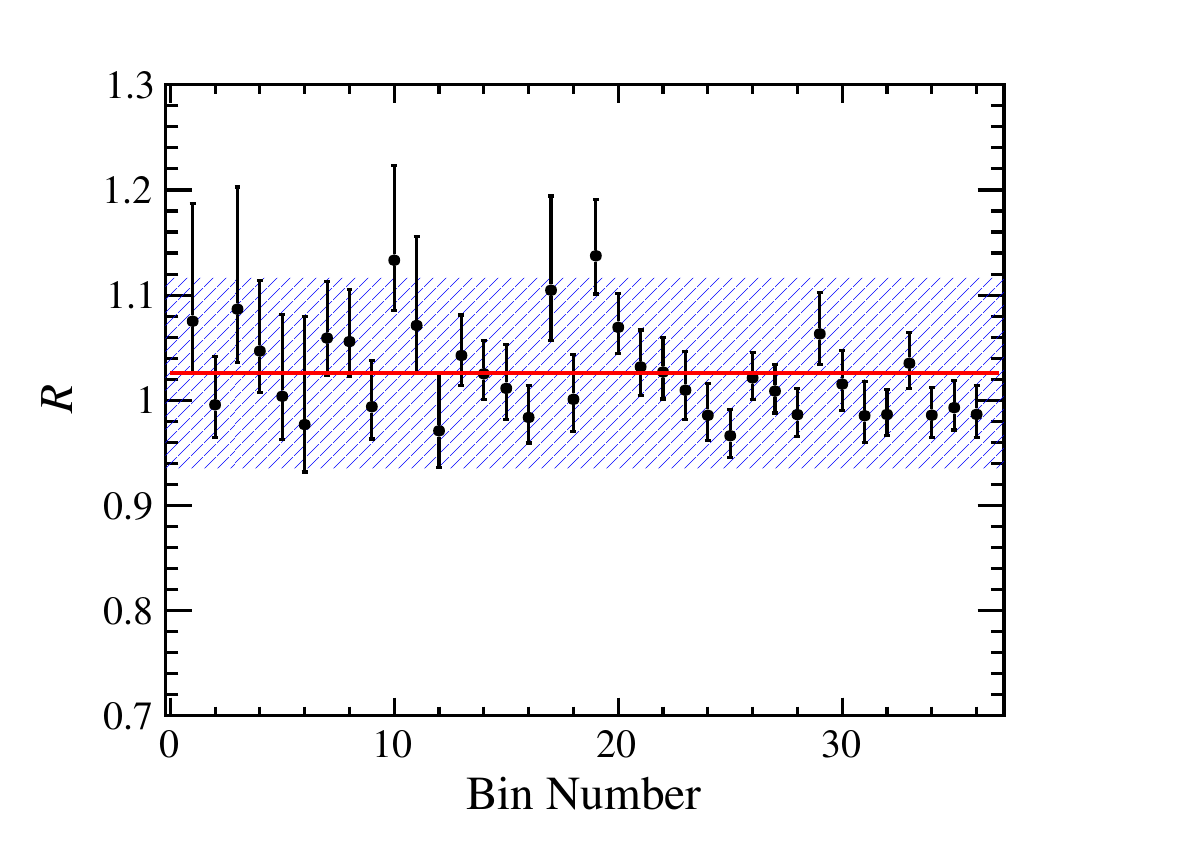}
    \caption{The ratio $R$ of measured slant depth ($X_\mathrm{meas}$) to expected slant depth ($X_\mathrm{exp}$) for each angular bin. The solid red line represents the mean value $\mu = 1.026$, and the blue lines indicate the $\mu \pm 2\sigma$ band. All data points are consistent with the mean.}
    \label{fig:ratio}
\end{figure}

Figure~\ref{fig:ratio} shows the distribution of $R(\theta,\varphi)$ for all measured directions in the data. We calculate the mean $\mu$ and standard deviation $\sigma$ of this distribution. Any direction where $R$ deviates from $\mu$ by more than $2\sigma$ is flagged as a potential anomaly. The resulting mean is $\mu = 1.026^{+0.009}_{-0.005}$ with a standard deviation of $\sigma = 0.045^{+0.028}_{-0.016}$, where the asymmetric uncertainties are propagated from data statistics only; systematic uncertainties from the terrain model, detector effective area, direction reconstruction, and rock composition are not included. The mean value of $1.026$ deviates from unity by approximately $3.5\sigma$ on the mean, suggesting the true average density of the mountain is about 2.6\% higher than the nominal value used in the calculation. This offset may also partly reflect uncertainty in the surface flux normalization; however, recent measurements suggest that post-LHC models tend to underestimate muon fluxes~\cite{KM3NeT:2024buf}, which would decrease rather than increase $R$, making a genuine density excess the more likely explanation. With 36 independent angular bins, the expected number of $2\sigma$ statistical false flags is approximately 1.6. All measured directions fall within the $\mu \pm 2\sigma$ band, consistent with no localized density anomalies within the probed volume, given our current statistical sensitivity.

In summary, this study not only reconstructs the mountain profile but also confirms the overall geological uniformity of Jinping Mountain. Alternatively, assuming that the measured density matches the nominal value used in the simulation, the results show no significant discrepancy with the \GEANT4 modeling of TeV-scale muon energy loss at the 95\% CL over slant depths from 2,400 to 3,600~m.w.e.

\section{Flux Prediction}\label{sec:prediction}
Underground muons represent a significant background for various physics experiments at CJPL, including searches for neutrinos and dark matter. Predicting the muon flux in each experimental hall is crucial for estimating muon-induced backgrounds and for related physics analyses. The total muon flux at an underground laboratory can be predicted as
\begin{equation}
    \phi_\mu = \int \phi_s(E,\theta, \varphi) P\left(E,X(\theta,\varphi)\right)dE d\Omega,
\end{equation}
where $\phi_s(E,\theta, \varphi)$ is the previously defined surface muon flux distribution and $P(E,X(\theta,\varphi))$ is the muon survival probability for the underground laboratory. This equation can be rewritten as
\begin{equation}
    \phi_\mu = \phi_0 \int f_s(E,\theta, \varphi) P\left(E,X(\theta,\varphi)\right) dE d\Omega = \phi_0 P,
    \label{eq:prediction}
\end{equation}
where $\phi_0$ is the total muon flux at the mountain surface, $f_s(E,\theta, \varphi)$ is the normalized surface muon distribution, and $P$ is the weighted survival probability according to $f_s(E,\theta, \varphi)$. In this work, we establish a \GEANT4-based simulation framework to calculate $P$ for a given underground laboratory and the surrounding mountain model~\cite{GEANT4:2002zbu,allison2006geant4}.

We use the NASA SRTM3 satellite dataset to model the mountain terrain within a 10~km radius of CJPL~\cite{terrain2007}. The Jinping mountain is modeled as uniform marble rock with a density of $2.8~\mathrm{g/cm}^3$~\cite{CDEX:2021cll,zheng2024three}. The mountain geometry is integrated into \GEANT4 using the Delaunay triangulation method. The position of CJPL-I relative to the mountain is determined using data described in Sec.~\ref{sec:location}, from which the positions of the CJPL-II experimental halls are derived. In the simulation, muons are generated on the mountain surface with energies and directions sampled from $f_s(E,\theta, \varphi)$. The weighted survival probability $P$ is then derived from the simulation results by counting the fraction of muons that reach the underground laboratories. Finally, the total muon flux in the underground laboratory can be predicted using Eq.~(\ref{eq:prediction}).

For a given underground laboratory, the predicted muon flux depends on the surface muon flux model, which affects both the weighted survival probability $P$ and the total surface flux $\phi_0$. As described in Section~\ref{sec:experiment}, the surface muon flux can be calculated theoretically based on hadronic interaction models. Variations in $P$ due to different hadronic models are estimated to be at the 1\% level, while the total surface muon flux $\phi_0$ can vary by up to 30\%, constituting the dominant impact on the predicted flux. Furthermore, several experiments have reported measured muon fluxes from the sub-TeV to multi-TeV range that are significantly higher than predictions from post-LHC hadronic models~\cite{KM3NeT:2024buf,Fedynitch:2021ima,Zhang:2025tmz}. Therefore, to provide an effective muon flux prediction for experiments at CJPL-II as a physics input, a relative method is employed, using the measured muon flux at CJPL-I as a reference:
\begin{equation}
    \phi_\mu = \frac{P_2}{P_1}\times\phi_{1},
\end{equation}
where $\phi_{1}$ is the muon flux at CJPL-I, taken from the measurement described in Sec.~\ref{sec:reconstruction}, and $P_1$ and $P_2$ are the weighted survival probabilities for CJPL-I and CJPL-II, respectively. This prediction method is largely independent of the chosen hadronic interaction and primary cosmic ray models.

Using the measured $\phi_1$ value as a reference, the total muon fluxes for each experimental hall at CJPL-II are predicted and presented in Table~\ref{tab:prediction}. Most of the experimental halls at CJPL-II have lower muon fluxes than the Sudbury Neutrino Observatory (SNO) laboratory, which previously reported the lowest muon flux before the construction of CJPL-II~\cite{SNO:2009oor}. The dominant uncertainty on $\phi_1$ is 4.7\%, originating from the measurement and including statistical uncertainty, as well as efficiency and effective area uncertainties. Further details are provided in Refs.~\cite{JNE:2020bwn,JNE:2024gov}.

\begin{figure}[!htbp]
    \centering
    \includegraphics[width=0.95\linewidth]{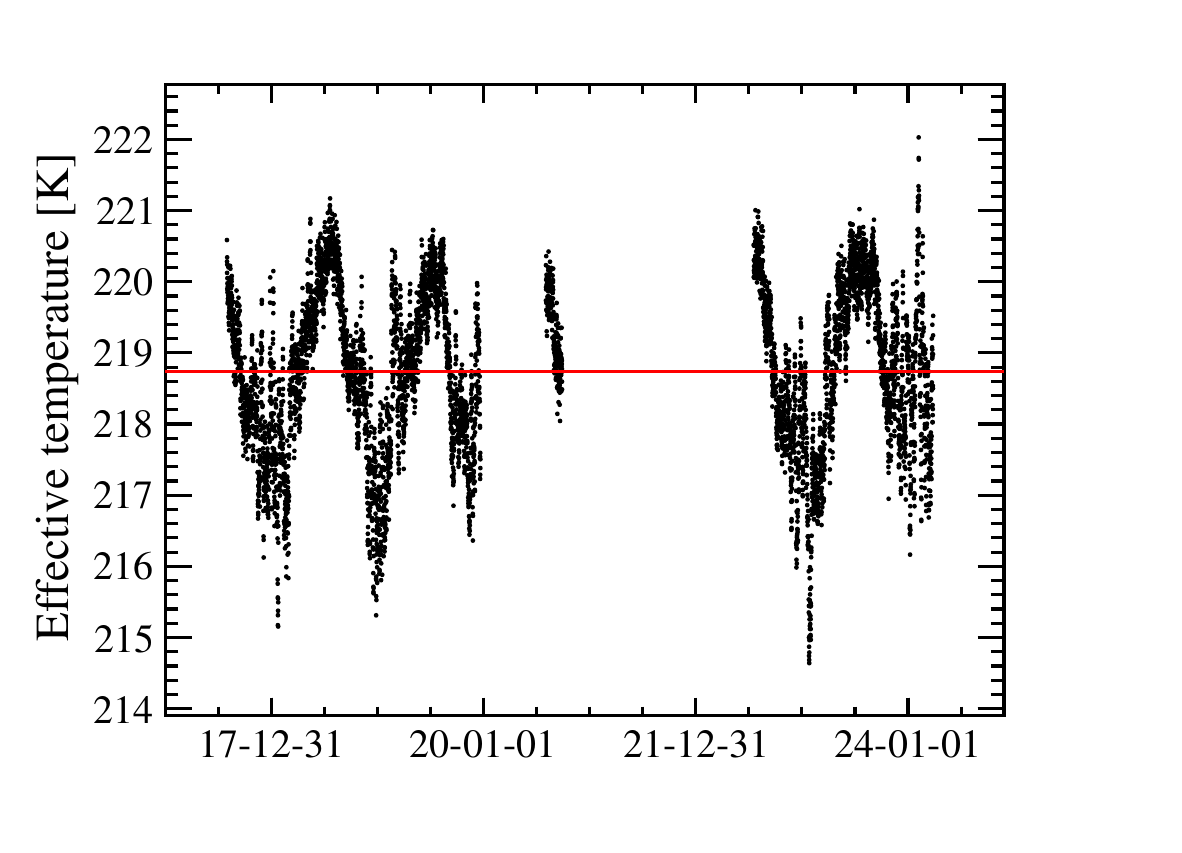}
    \caption{Calculated time-varying effective temperature, $T_\mathrm{eff}$, during the data-taking period. The solid points show the daily calculated $T_\mathrm{eff}$, and the red line indicates the mean value over the entire period.}
    \label{fig:season}
\end{figure}

Moreover, measured muon fluxes are affected by atmospheric conditions, especially atmospheric temperature and density~\cite{Gaisser:2016uoy}. Given that our data-taking period is not distributed evenly throughout the year, there is a potential bias in the $\phi_1$ measurement. To account for this effect, we retrieved temperature data from the ERA-Interim reanalysis of the European Centre for Medium-Range Weather Forecasts during the data-taking period around CJPL~\cite{ERA52011}. Using the theoretical framework presented in Ref.~\cite{Grashorn:2009ey} and parameters from Ref.~\cite{MINOS:2009njg}, we calculated the conversion coefficient linking temperature fluctuations to muon flux variations at CJPL-I to be 0.95. The time-varying effective temperature, $T_\mathrm{eff}$, is determined by weighting temperatures at different atmospheric levels (Fig.~\ref{fig:season}). The standard deviation of $T_\mathrm{eff}$ over the data-taking period is $\approx 1.1$~K, corresponding to an estimated uncertainty of 0.5\% in $\phi_1$.

For the $P_2 / P_1$ ratio, the uncertainties primarily arise from the simulation of the mountain overburden. We study the induced uncertainties by varying the surface muon flux models and the position of CJPL-I in the mountain, while keeping the relative position between CJPL-I and CJPL-II fixed. The uncertainty in $P_2 / P_1$ induced by different surface flux models is estimated at 0.7\%. The CJPL-I position is varied by $\pm 30$~m around its nominal position in the simulation. This variation induces uncertainty in $P_2 / P_1$, determined to be 0.9\%. Additionally, the statistical uncertainty from the MC simulation is less than 0.4\%, which is negligible compared to other sources. Finally, the total uncertainty in $P_2 / P_1$ is estimated at 1.1\%, indicating that the dominant source of uncertainty in the flux prediction is the flux measurement at CJPL-I.

In Ref.~\cite{zhang2025measurement}, the muon flux in the C2 hall was measured by a plastic scintillator system to be $(3.03 \pm 0.24_{\mathrm{stat.}} \pm 0.18_{\mathrm{syst.}})\times10^{-10}~\mathrm{cm}^{-2}\mathrm{s^{-1}}$, which is consistent with the prediction in this work. We also derived the average underground muon energy at CJPL ($E_{\mu}$) based on the mountain simulation and the SIBYLL-2.3d hadronic model, as presented in Table~\ref{tab:prediction}. This parameter directly affects muon-induced neutron production in the surrounding rock and detector components, a dominant background source for both neutrino and dark matter experiments~\cite{Mei:2005gm}. In Ref.~\cite{Woodley:2024eln}, the value of $E_{\mu}$ at CJPL-I was also predicted using a different method from this study, yielding a consistent result. In a previous study~\cite{JNE:2024gov}, the Jinping Mountain rock was assumed to have the composition of the Earth's crust, in which case the muon energy loss is lower than in marble. This assumption led to a higher calculated $E_{\mu}$ for CJPL-I than in the present work.

\begin{table}[!htbp]
    \tabcolsep=0.2cm
    \centering
    \caption{The total muon fluxes in the different experimental halls of CJPL predicted with the relative method. The muon flux at CJPL-I is the measured value. The uncertainties here include the statistical and systematic uncertainties in the measurement, and the systematic uncertainties from the simulation. Details are discussed in the text. The average underground muon energies $E_{\mu}$ for each hall are also listed, predicted based on the mountain simulation and the SIBYLL-2.3d hadronic model.}\label{tab:prediction}
    \begin{tabular}{ccc}
        \hline
        \hline
        Experiment Hall &  Muon Flux~[$10^{-10}~\mathrm{cm}^{-2}\mathrm{s^{-1}}$] & $E_{\mu}$~[GeV] \\
        \hline
         CJPL-I &  3.54$\pm$0.17 (Measurement)    & 328     \\
         ~& ~& ~\\
         CJPL-II &     &       \\
         A1 (JUNA) &  3.28$\pm$0.16    &  327    \\
         A2 &  3.45$\pm$0.17    &   341    \\
         B1 &  3.18$\pm$0.15    &   333    \\
         B2 (PandaX) &  3.23$\pm$0.16    &  358    \\
         C1 (CDEX) &  2.85$\pm$0.14    &  339    \\
         C2 &  2.89$\pm$0.14    &   341    \\
         D1 &  2.82$\pm$0.14    & 332     \\
         D2 (JNE) &  2.80$\pm$0.14  &  334     \\
        \hline
        \hline
    \end{tabular}
\end{table}

\section{Conclusions and Outlook}\label{sec:conclusion}
In this study, we present a detailed muography study of Jinping Mountain, based on 1338.6 live days of data from the one-ton prototype of JNE. The energy and direction of collected muon signals are reconstructed, achieving angular reconstruction accuracy of approximately $4.5^\circ$ and consistent detection over the $4\pi$ solid angle. Muon slant depths in each direction are derived based on the calculated surface fluxes and the measured underground flux. These slant depths are then used to reconstruct the mountain structure around CJPL-I within a 3~km radius. The reconstructed structure is compared with satellite data, with all directions agreeing within $2\sigma$, indicating no significant density variations within the probed volume, given our current statistical sensitivity. Based on the muography results and the measured flux at CJPL-I, we performed a relative prediction of the total muon fluxes for the eight halls in CJPL-II, providing an essential input for future physics research at these facilities. The predicted results exhibit consistency with available measurements and other studies. We estimate that this muography method is effectively applicable to imaging density anomalies of 100~m.w.e.\ scale in mountains with heights up to 600~m. A seasonal correction to the measured flux, based on atmospheric temperature variations, contributes an additional 0.5\% systematic uncertainty.

\begin{acknowledgments}
This work was supported in part by the National Natural Science Foundation of China (12127808, 12141503, 12305117, 12441513, and 12521007), the Key Laboratory of Particle and Radiation Imaging (Tsinghua University), and the State Key Research Development Program in China (Nos. 2022YFA1604700). We acknowledge Orrin Science Technology, Jingyifan Co., Ltd, and Donchamp Acrylic Co., Ltd, for their efforts in the engineering design and fabrication of the stainless steel and acrylic vessels. Many thanks to the CJPL administration and the Yalong River Hydropower Development Co., Ltd. for logistics and support.
\end{acknowledgments}

\bibliographystyle{apsrev4-2}
\bibliography{bibfile}

\end{document}